\RequirePackage{fix-cm}
\documentclass{svjour3}                         
\smartqed  
\usepackage{graphicx}
\usepackage{amssymb}
\usepackage{amsmath}
\usepackage{graphicx}
\usepackage{txfonts}
\usepackage{booktabs}
\usepackage{textcomp}
\usepackage{color}
\usepackage[caption=false]{subfig}
\usepackage{url}
\usepackage{xcolor}
\usepackage{hyperref}
\hypersetup{colorlinks,urlcolor=blue}
\begin{document}

\title{Methods for detection and analysis of weak radio sources with single-dish radio telescopes}

\titlerunning{Detection methods for radio faint sources}        

\author{M.~Marongiu         \and
        A.~Pellizzoni       \and
        E.~Egron            \and
        T.~Laskar           \and
        M.~Giroletti        \and
        S.~Loru             \and
        A.~Melis            \and
        G.~Carboni          \and
        C.~Guidorzi         \and
        S.~Kobayashi        \and
        N.~Jordana-Mitjans  \and
        A.~Rossi            \and
        C.~G.~Mundell       \and
        R.~Concu            \and
        R.~Martone          \and
        L.~Nicastro}

\authorrunning{M.~Marongiu et al.} 

\institute{M.~Marongiu \and R.~Martone
\at Department of Physics and Earth Science, University of Ferrara, via Saragat 1, I--44122 Ferrara, Italy
\at ICRANet, Piazzale della Repubblica 10, I--65122 Pescara, Italy \\
\email{marco.marongiu@unife.it}
\and A.~Pellizzoni \and E.~Egron \and A.~Melis \and G.~Carboni \and R.~Concu
\at INAF, Osservatorio Astronomico di Cagliari, Via della Scienza 5, I--09047 Selargius, Italy \\
\and M.~Giroletti
\at INAF, Istituto di Radioastronomia, Via Piero Gobetti 101, I--40129 Bologna, Italy \\
\and S.~Loru
\at INAF--Osservatorio Astrofisico di Catania, Via Santa Sofia 78, I--95123 Catania, Italy \\
\and T.~Laskar \and N.~Jordana-Mitjans \and C.~G.~Mundell
\at Department of Physics, University of Bath, Claverton Down, Bath, BA2 7AY, UK \\
\and C.~Guidorzi
\at Department of Physics and Earth Science, University of Ferrara, via Saragat 1, I--44122 Ferrara, Italy \\
\and S.~Kobayashi
\at Astrophysics Research Institute, Liverpool John Moores University, Liverpool, L3 5RF, UK \\
\and A.~Rossi \and L.~Nicastro
\at INAF--Osservatorio di Astrofisica e Scienza dello Spazio di Bologna, Via Piero Gobetti 93/3, I--40129 Bologna, Italy \\
}

\date{Received: 30 December 2019 / Accepted: 25 March 2020}


\maketitle

\begin{abstract}
The detection of mJy/sub-mJy point sources is a significant challenge for single-dish radio telescopes.
Detection or upper limits on the faint afterglow from GRBs or other sources at cosmological distances are important means of constraining the source modeling.

Using the Sardinia Radio Telescope (SRT), we compare the sensitivity and robustness of three methods applied to the detection of faint radio sources from raster maps around a known source position: the smart ’quick-look’ method, the ’source extraction’ method (typical of high-energy astronomy), and the fit with a 2-D Gaussian.
We developed a Python code specific for the analysis of point-like radio sources applied to the SRT C-band ($6.9$~GHz) observations of both undetected sources (GRB afterglows of 181201A and 190114C) and the detected Galactic X-ray binary GRS\,1915+105.

Our comparative analysis of the different detection methods made extensive use of simulations as a useful complement to actual radio observations.
The best method for the SRT data analysis is the fit with a 2-D Gaussian, as it pushes down the sensitivity limits of single-dish observations -- with respect to more traditional techniques -- to $\sim 1.8$~mJy, improving by $\sim 40$\,\% compared with the initial value.
This analysis shows that -- especially for faint sources -- good maps of the scanned region pre- or post-outburst are essential.

\end{abstract}

\section{Introduction}
\label{sec:intro}
The detection of faint (mJy/sub-mJy) point sources with single-dish radio telescopes require (1) an accurate knowledge of the background (both instrumental and astronomical), (2) good sky opacity conditions to ensure an accurate calibration, and (3) a reliable and well-defined source detection method.
Whenever the source is not detected, it is common practice to estimate an upper limit based on the flux density root mean square (RMS) calculated over the image\footnote{The root mean square (RMS) of the image is the standard deviation of pixels flux densities taken in image regions not affected by the source flux.}.

With this aim, among several detection methods for the analysis of sources in single-dish mode (e.g., \cite{Giroletti20,Maddalena02}), we examine three methods using the network of radio telescopes of the National Institute for Astrophysics (INAF), which includes the Sardinia Radio Telescope (SRT), the Medicina Radio Astronomical Station, and Noto Radio Observatory\footnote{\url{http://www.radiotelescopes.inaf.it/}}.
These are 'quick-look' (Method A, a smart but rough approach), 'source extraction' (Method B, typically adopted in X-ray/gamma-ray astronomy), and fitting procedure with a 2-D Gaussian (Method C, a more sophisticated approach accounting for the instrument point spread function).

The science case of study is GRB radio afterglows, a phenomenon associated with very faint sources (with flux densities $\lesssim 1$~mJy), and hence, very suitable to our analysis (e.g. \cite{ChandraFrail12,Chandra16}).
GRBs are detected through their bright and characteristic gamma-ray prompt emission and more recently, as the counterparts of gravitational waves (GWs; e.g. \cite{Abbott17,Alexander17gw,Margutti18a}).
GRB ejecta produce a relativistic blast wave shock as they expand into their ambient environment.
This shock accelerates electrons and produces synchrotron radiation, which is visible as long-lasting X-ray to radio ``afterglow'' emission.
Observations of GRB afterglows at radio frequencies provide a wealth of information: (1) constrain the self-absorption frequency of the underlying synchrotron radiation \cite{GranotSari02}, and thus break parameter degeneracies in conjunction with optical and X-ray observations, (2) track the presence and evolution of reverse shocks in the ejecta and hence derive the ejecta magnetization and initial Lorentz factor \cite{Kulkarni99,Laskar13,Laskar19a}, (3) constrain the degree of ejecta collimation and hence the released energy corrected for relativistic beaming \cite{Rhoads99,Sari99b}, and (4) derive the size of the afterglow using scintillation methods \cite{Frail00,Alexander19}.
Overall, they contribute remarkably to our understanding of the hydrodynamics of relativistic outflows.

We analyzed the performance of the three detection methods through dedicated radio followup campaigns of two GRB afterglows (GRB\,181201A and GRB\,190114C) in C-band ($6.9$~GHz) with SRT, resulting in upper limits.
The information on the position of these sources comes from the detection of afterglow counterparts both at X-ray and optical wavelengths.
We make extensive use of simulations of point-like sources, injected in simulated images/fields to quantify our upper limit errors.

This paper is organized as follows.
We describe our targets in Section~\ref{sec:target}, our radio observations in Section~\ref{sec:observ}, whereas Section~\ref{sec:imagingcalib} explains the imaging data analysis and the calibration procedure in single-dish mode.
Following the description of the three detection methods for point-like sources (Sect.~\ref{sec:detmethods}), in Section~\ref{sec:sim} we apply them to a simulated case (background and source).
These simulations are crucial to analyze the real cases of undetected GRBs, and the faint source GRS\,1915+105 (Sect.~\ref{sec:reale_tutto}).
We present our results in Section~\ref{sec:discussion} and our conclusions in Section~\ref{sec:conc}.

\section{Our targets}
\label{sec:target}

We observed the fields of two long GRBs (181201A and 190114C) and the accreting black hole X-ray binary GRS\,1915+105 with SRT in C-band ($6.9$~GHz).
Even hours after the burst, the optical afterglows of GRB\,181201A and GRB\,190114C optical afterglow remained bright (magnitude $< 18$ in R-filter; e.g. \cite{Volnova18} for GRB\,181201A and \cite{Izzo19b} for GRB\,190114C).

GRB\,181201A was discovered by the INTEGRAL Burst Alert System (IBAS) in IBIS / ISGRI data on 2018 December 1 at 02:38 UT \cite{Mereghetti18}; it was also detected by the Fermi Large Area Telescope (LAT; \cite{Arimoto18}), the X-Ray Telescope (XRT) on Swift \cite{Pintore18} and the High-Energy (HE) instrument aboard Insight-HXMT \cite{Cai18}.
The afterglow was observed in optical (e.g. \cite{Watson18}), and mm/radio frequencies \cite{Laskar19b}, with possible evidence for an associated supernova \cite{Belkin19}.
It has a redshift of $0.450$ \cite{Izzo18}.
We centred our observations of GRB\,181201A on the Karl G. Jansky Very Large Array (VLA) coordinates $\alpha=21^{\rm h}17^{\rm m}11^{\rm s}.185$ and $\delta=-12^{\circ}37'51.37''$, refined by optical/X-ray observations \cite{Laskar18b}.

GRB\,190114C was discovered by the Burst Alert Telescope (BAT; \cite{Barthelmy05}) on the Neil Gehrels Swift Observatory \cite{Gehrels04} on 2019 January 14 at 20:57:03 UT \cite{Gropp19}.
With a redshift $z = 0.4245$ \cite{Castro-Tirado19}, it was also detected by Konus-Wind \cite{Frederiks19}, the Fermi Gamma-ray Burst Monitor (GBM; \cite{Meegan09}), the Fermi/LAT (\cite{Hamburg19,Kocevski19}), and radio facilities (e.g. ALMA, \cite{Laskar19a}; ATCA, \cite{Misra19}).
GRB\,190114C is the first GRB detected in the TeV band ($\ge 300$~GeV) by the twin Major Atmospheric Gamma Imaging Cherenkov (MAGIC) telescopes, with observations starting $50$~s after the BAT trigger \cite{Mirzoyan19,Mirzoyan19b,MAGIC19b}.
For our observations of GRB\,190114C, we used the VLA coordinates $\alpha=3^{\rm h}38^{\rm m}01^{\rm s}.191$, $\delta=-26^{\circ}56'46.73''$, refined by optical/X-ray observations \cite{Alexander19}.

GRS\,1915+105 is a highly variable accreting black hole X-ray binary in our Galaxy.
In the radio it shows relativistic superluminal jets of flux density $\sim 1$\,Jy \cite{Mirabel94,Fender99} and compact jets with flux density of $20-200$\,mJy \cite{Pooley97,Eikenberry98}.
In the framework of the monitoring of GRS\,1915+105 with SRT (PI: Egron, proposal 28-18), we performed observations on 22 May 2019 at $6.9$~GHz when the source was slightly active.
We carried out the data analysis with our software in order to test our methods of detection on a quite weak but clearly detected source with SRT and compare the result of the flux density with another software dedicated to single dish imager observations (SDI, \cite{Egron17b}).

\section{Observations with SRT}
\label{sec:observ}

\begin{figure*} 
\centering
{\includegraphics[width=58mm]{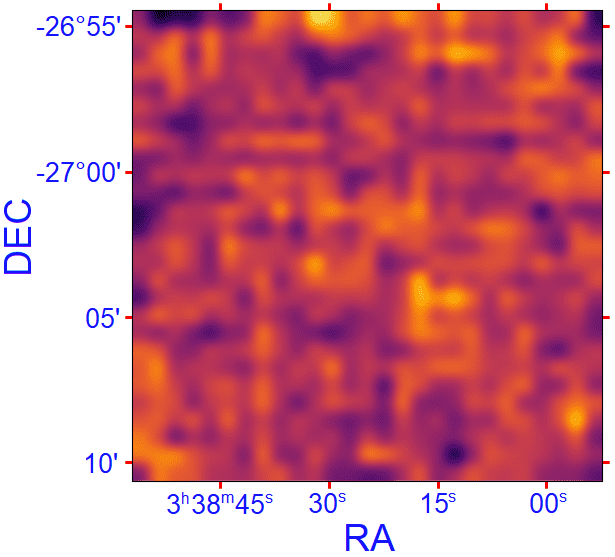}} \quad
{\includegraphics[width=95mm]{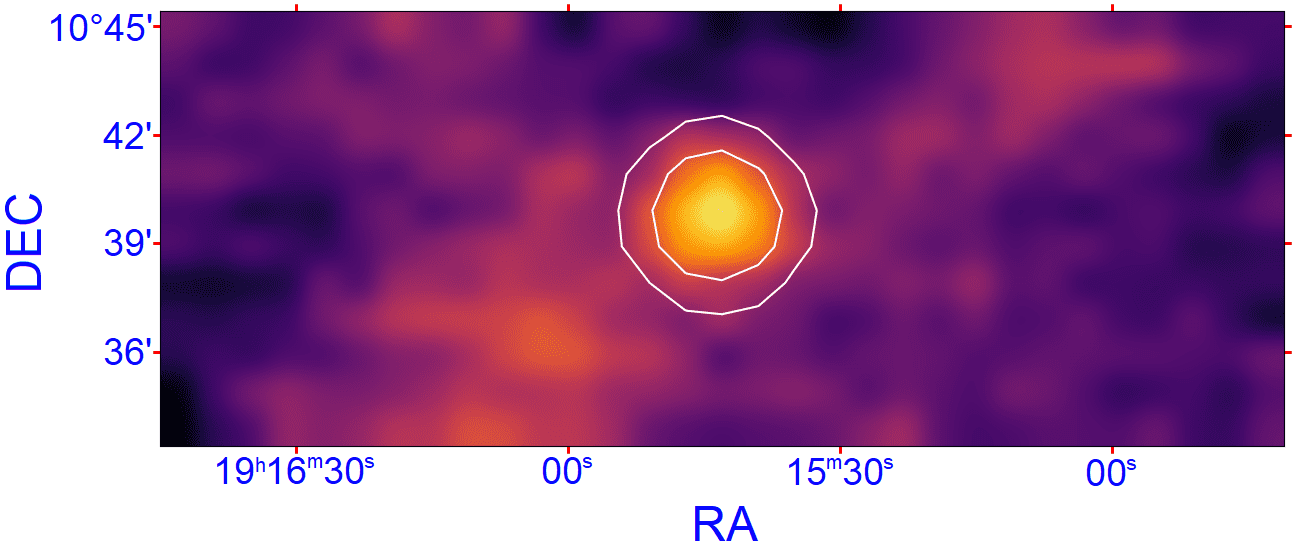}} \quad
\caption{
Gaussian smoothed images of GRB\,190114C (top, observed on 2019 January 17), and GRS\,1915+105 (bottom).
The features seen in the image of GRB\,190114C can be ascribed to RFI.
White circles indicate contours with respect to $2$ (outer) and $5$ (inner) times the value of $RMS_{min}$.
}
\label{fig:im_real}
\end{figure*}

We observed GRB\,181201A, GRB\,190114C and GRS\,1915+105 with SRT\footnote{\url{http://www.srt.inaf.it/}} (Fig.~\ref{fig:im_real}), which is part of the INAF radio telescope network; it is the largest among them with a 64-m diameter parabolic reflector.
SRT is located in San Basilio (near Cagliari, Sardinia, Italy) and designed to operate in the $0.3 - 100$~GHz nominal frequency range with a Gregorian configuration \cite{Prandoni17}.
At present, SRT is equipped with three receivers: a coaxial dual-feed L--P-band ($1.3 - 1.8$~GHz; $305 - 410$~MHz) receiver, a mono-feed C-band receiver ($5.7-7.7$~GHz), and a 7-beam K-band receiver ($18 - 26.5$~GHz) \cite{Valente10,Valente16}.
An active surface (composed of $1008$ panels and $1116$ electro-mechanical actuators) implemented on the primary mirror allow us to (1) compensate for gravitational deformations and (2) re-shape the primary mirror from a shaped configuration to a parabolic profile \cite{Orfei04,Buttu12,Bolli15,Egron17}.
The targets were observed with SRT at $6.9$~GHz (bandwidth = $1200$~MHz) between 2018 December 11 and 2019 March 22 under the project 23-18 for GRB\,181201A (PI: M.~Marongiu) and the ToO request 02-19 for GRB\,190114C (PI: M.~Marongiu).
We observed in ``shared-risk mode'' through a new-generation and flexible ROACH2-based backend called {\sc sardara} (SArdinia Roach2-based Digital Architecture for Radio Astronomy \cite{Melis18}).

During {\sc sardara} operations, the total bandwidth of each polarization (LCP and RCP) is divided into $1024$ channels.
For continuum observations, this allows us to dynamically remove radio frequency interference (RFI) and thus maximize our point-source sensitivity.
\begin{table*}
\centering
\caption{Observation campaign for our analysis with SRT.
$\Delta t_i$ and $\Delta t_f$ indicate respectively the start and final observing epoch after the GRB explosion (in units of days).}
\label{tab:obstot}
\begin{tabular}{|ccc|cc|}
\hline
Epoch & GRB & Receiver & $\Delta t_i$ & $\Delta t_f$ \\
 & & & (d) & (d) \\
\hline \hline
2018/12/11 & 181201A & SRT-C & 10.36 & 10.52 \\
2019/01/30 & 181201A & SRT-C & 60.27 & 60.49 \\
2019/03/22 & 181201A & SRT-C & 111.12 & 111.32 \\
\hline
2019/01/17 & 190114C & SRT-C & 2.77 & 2.99 \\
2019/01/23 & 190114C & SRT-C & 8.74 & 8.94 \\
2019/03/05 & 190114C & SRT-C & 49.66 & 49.84 \\
\hline
\end{tabular}
\end{table*}
We performed the mapping of our targets through On-the-Fly (OTF) scans.
Within this technique, data are continuously stored while the antenna performs constant-speed orthogonal scans across the sky, alternately producing maps along the Right Ascension (RA) and Declination (Dec) directions. Unlike raster maps (where separated on-source/target and off-source/background pointing are performed), in OTF mapping the target signal is measured together with the background/baseline level (continuously spanning a sky region larger than the target).
This allows for a more precise background/baseline subtraction.
Observations were carried out through the repetition for $\sim 15$ RA/Dec maps of $0.2^{\circ} \times 0.2^{\circ}$ with $4$~arcmin/s scan speed, $4.5$~scan/beam, and a sampling time of $20$~ms during the observations.
The dimensions of the maps were chosen based on the beam size in C-band at the observing frequency $\nu_{obs} = 6.9$~GHz ($HPBW = 2.71 \pm 0.02$~arcmin \cite{Bolli15})\footnote{The half power beamwidth (HPBW) is the angular separation in which the magnitude of the radiation pattern decrease by $50 \%$ (or $3$~dB) from the peak of the main beam.}.
This configuration allows us to reach an exposure of $\sim 1$ minute/beam for a total mapping time of $\sim 6$~hours including overheads and slew time, in order to have the chance to detect a counterpart at $\sim$mJy level or below.
This observing strategy provides a direct image of the sources close to the target as well as a better estimate of the flux density.

An accurate evaluation of flux density errors is possible thanks to the acquisition of $> 10-20$~samples/beam for each scan passage; this generates a large beam oversampling (with respect to Nyquist sampling), that allows us to efficiently remove outlier measurements ascribed to RFI.
The length of the scans is chosen based on the source size; the scan-dependent baseline (i.e. background emission and system-related signal) must be correctly subtracted, to properly reconstruct the morphology of the observed source and its associated flux density \cite{Egron17}.
Ideally, each scan should be free of significant source contribution (and RFI contamination) for $40-60$ per cent of its length/duration, to properly identify and subtract the baseline component; usually this requirement is satisfied for extragalactic targets (i.e. GRBs), but not trivially satisfied for targets located in crowded regions of the Galactic plane.

Two consecutive scans were separated by an offset of exactly $0.01^{\circ}$, which implies that -- assuming a beam size of $2.7$~arcmin in C-band -- (1) $4.5$ passages were carried out per beam on average, and (2) $\sim 17$~samples beam$^{-1}$ scan$^{-1}$ were taken.
The total duration of an observation (defined as a complete map along both RA and Dec directions) at $6.9$~GHz was about $6$~h.
Stable weather conditions (possibly a clear sky) result in the production of high-quality maps; in only two epochs (2019 January 17 and 2019 March 5) weather conditions were excellent during the observation (Table~\ref{tab:obstot}), while in the other epochs high cloud coverage and rainy conditions provided poor and variable opacity. 

The spectral flux density of the target was reconstructed by performing a set of OTF cross-scans on standard point-like calibrators at the relevant frequencies (3C286, 3C295, 3C123, 3C48, 3C147 and NGC7027) before and after each target map, assuming calibrator fluxes as obtained by \cite{Perley13d}, using the VLA data \cite{Egron16}.

\subsection{Imaging data analysis and calibration}
\label{sec:imagingcalib}

The imaging procedure is performed through SRT Single-Dish Imager ({\sc SDI}), a tool optimized for OTF scan mapping; {\sc SDI} performs automated baseline subtraction, RFI rejection and calibration, generating standard SAOImage DS9\footnote{\url{http://ds9.si.edu}} output FITS images suitable for further analysis \cite{Egron16,Egron17}.

The conversion factor Jy/counts ($K_{conv}$) for the calibration is defined as the ratio between the peak flux density $S_{cal}$ (at that specific observing frequency) and the maximum value of the observed instrumental counts in the calibrator image $C_{max}$.
This factor is roughly independent of the elevation since C-band SRT gain curve is approximately flat (within a few percent) thanks to optimized settings of the active surface (shaped mode) \cite{Bolli15,Egron16,Prandoni17,Egron17}.
We considered calibrators and target observations in the elevation range $\sim 30$--$80^{\circ}$ since the antenna beam has proved to be very stable in that interval.
A Gaussian shape provides a very good fit to OTF scans on calibrators, thus we assumed a beam solid angle for image calibration (in units of steradians) as \cite{Egron17}:
\begin{equation}
\Omega_{beam} = \pi \left(1.2 \times \frac{HPBW}{2}\right)^2 \;,
\label{eq:solanglebeam}
\end{equation}
where $HPBW$ is in units of arcmin.

Calibrated data were binned through ARC-tangent projection using pixel sizes $\sim 1/4$ of the HPBW, corresponding to the effective resolution of the images \cite{Egron17}, and the FITS images were produced in units of Jy/beam and Jy/sr.
The statistical errors on flux density measurements are calculated through the flux density standard deviation for each pixel; moreover, the integrated statistical flux errors (typically $< 0.5$\%) are well below the systematic errors, estimated to be $\lesssim 3$\% \cite{Egron17}.

\section{Detection methods for point sources}
\label{sec:detmethods}

Our study tackles the question ``How is optimally detected a weak source surrounded by other sources and affected by background?'' To answer it, we need a robust, reliable and sensitive detection method.

We analyze three detection methods according to sensitivity and robustness: `quick-look' (Method A, Sect.~\ref{sec:quick}), `source extraction' (Method B, Sect.~\ref{sec:he_detect}), and fitting procedure with a 2-dimensional Gaussian (Method C, Sect.~\ref{sec:2D_gauss}).
These methods are applicable to radio detection, such as the case of highly variable sources, which can have a very weak radio flux density during periods of quiescent states, and/or in a crowded regions of the sky.
This is the case for instance of the microquasar GRS\,1915+105.
Regarding undetected (or very faint) sources, in the radio domain the upper limit for the flux density is usually estimated as twice the minimum RMS of a region in the image not significantly affected by other sources ($RMS_{min}$).
Such upper limits can be overestimated if the target is surrounded by other sources (crowded field), and in any case it does not represent the actual sensitivity at the source position.

For this analysis we developed a specific Python code, where the input is the calibrated (in units of flux density/beam) FITS image (suited for INAF network) produced by {\sc SDI}, and the output consists of the flux density $A_m$ and its uncertainty $\Delta A_m$.
This code will be directly implemented in the SDI package soon.
We define the significance as the signal-to-noise ratio $\mathcal{S} = A_{m} / \Delta A_{m}$, using each technique (in turn characterized by different uncertainties).
Table~\ref{tab:quickand2} reports the upper limits obtained with each method.
It is worth noting that our analysis does not consider the systematic errors of the antenna, as they are negligible compared with the statistical ones for very weak sources (Sect.~\ref{sec:imagingcalib}).

\subsection{Method A - Quick-look detection method}
\label{sec:quick}

Method A consists of an estimation of the flux density of the target $S_{source}$, corresponding to the pixel of the source position (peak flux of the point-like target taken in Jy/beam units); the uncertainty is defined as $RMS_{min}$, the minimum RMS of a fixed region (a rectangle of size $1.5$~beam for this work) in the image.
In case the source is not detected, this is a suitable method to get a rough upper limit; we assume $2 \times RMS_{min}$ as upper limit.

\subsection{Method B - Source extraction method}
\label{sec:he_detect}

Method B is typical of X-/gamma-ray and optical imaging.
At the beginning, the procedure extracts the total flux in two regions:
\begin{enumerate}
\item the region --centered on the source-- with a radius of $n$~HPBW, depending on the maximum $S/N$ ratio in the range $1.5$ -- $5$~HPBW ($T_{s,sum}$); the $n$ value depends on the source and local background flux densities.
\item a background area taken from an image region free of sources ($T_{b,sum}$).
\end{enumerate}

After normalizing $T_{s,sum}$ and $T_{b,sum}$ with respect to the same extraction area, the contribution of residual background from the image is calculated through the difference $D = T_{s,sum} - T_{b,sum}$.
The flux density of our target (in units of Jy) is calculated as $S = D \times K_{conv} \times P$, where $K_{conv}$ is the calibration factor (Sect.~\ref{sec:imagingcalib}) and $P = \Omega_{pix}/\Omega_{beam}$ (where $\Omega_{pix}$ is the solid angle of a single pixel of the image).
The uncertainty on $S$ is calculated as $\Delta S = \Delta T_{s,sum} + \Delta T_{b,sum}$; the uncertainties $\Delta T$ correspond to $\sqrt{N_{pix}} \times RMS_{min}$, where $N_{pix}$ is the total number of pixels of the extraction region.
\begin{table*}
\centering
\scriptsize
\caption{Flux densities for our three sources using methods A, B and C (fixed position and position criterion) with SRT in C-band.
In case of non-detection (numbers in bold font), we injected a fake sources in the image, and the target is detected at $3\sigma$-level; therefore we assumed the upper limit as $2\sigma$-level of the $3\sigma$-level detection.}
\label{tab:quickand2}
\begin{tabular}{cc|cccc}
\hline
Epoch & Source & Method A & Method B & Method C (fixed) & Method C (free) \\
(aaaa/mm/dd) & & (mJy) & (mJy) & (mJy) & (mJy) \\
\hline
2018/12/11 & GRB\,181201A  & \textbf{3.9}    & \textbf{2.2}    & \textbf{7.4}    & \textbf{4.6}   \\
2019/01/17 & GRB\,190114C  & \textbf{3.4}    & \textbf{2.5}    & \textbf{1.8}    & \textbf{2.7}   \\
2019/01/23 & GRB\,190114C  & \textbf{6.5}    & \textbf{5.1}    & \textbf{3.0}    & \textbf{6.5}   \\
2019/01/30 & GRB\,181201A  & \textbf{11.0}   & \textbf{8.5}    & \textbf{7.5}    & \textbf{7.6}   \\
2019/03/05 & GRB\,190114C  & \textbf{5.2}    & \textbf{2.4}    & \textbf{2.5}    & \textbf{3.5}   \\
2019/03/22 & GRB\,181201A  & \textbf{36.4}   & \textbf{25.9}   & \textbf{18.2}   & \textbf{24.0}  \\
2019/05/22 & GRS\,1915+105 & $159.0 \pm 14.7$ & $177.0 \pm 11.7$ & $187.0 \pm 8.9$  & - \\
\hline \hline
\end{tabular}
\end{table*}

\subsection{Method C - 2-Dimensional Gaussian fitting}
\label{sec:2D_gauss}

In Method C we fit the image using a 2-D Gaussian on a flat background, defined as: 
\begin{equation}
G =\ N_0 + A \times e^{-u(x,y,x_0,y_0,\theta,a,b)/2}\ \;,
\label{eq:gauss2d}
\end{equation}
where $N_0$ is the residual background, $A$ is the amplitude and $u$ is the ellipse equation, dependent on the position ($x_0$, $y_0$), the semimajor and semiminor axes ($a$, $b$) and the position angle $\theta$\footnote{For our purpose, we fixed $a = b$ (circular beam) and $\theta=0$.}.

Our Python code adopts a non-linear least squares method through the Python package {\sc curvefit}.
The parameter uncertainties are described by the square root of the diagonal elements of the covariance matrix.

\section{Setting up of the code: simulations of point-like sources}
\label{sec:sim}

To test the robustness of our fitting procedure, we inject a fake point-like source in image convolved with a 2-D Gaussian (Eq.~\ref{eq:gauss2d}) with fixed HPBW ($a = b = 2.7$~arcmin, corresponding to the SRT C-band beam).
We then create a sample of images, each of which composed of one fake point-like source with increasing Gaussian peak $A$ in order to understand when the target becomes distinguishable from the background. In this way we tested the robustness of our fitting procedure by injecting fake point sources (one per image per trial) of varying amplitudes. We implemented two kinds of simulations: (1) full simulation of the source and the background (Sect.~\ref{sec:totsim}), and (2) simulation of the source in a real background/image (Sect.~\ref{sec:realmaps}).
Finally, for each image of this sample we apply these detection methods, comparing them in terms of sensitivity and robustness.
For Method C, we assume $A_m$ as free parameter, and we apply two combinations to localize the target in the image during the fitting procedure: (1) we fix the target position parameters ($x_{0,m}$, $y_{0,m}$) to their true values, which are known in the simulation (fixed position case), and (2) we assume $x_{0,m}$ and $y_{0,m}$ as free parameters (free position case).
Regarding the free position case, we assume two additional detection criteria: (1) the positional uncertainty $\Delta x$ and $\Delta y$ (obtained by fitting procedure) must be $\lesssim 10 \%$ of the actual source position, and (2) $x_{0,m}$ and $y_{0,m}$ must to be inside the $4 \%$ region of the true position. If these conditions are not satisfied, we consider it a non-detection.

We assume that -- for each detection method  -- the source is detected at $3\sigma$-level, whereas upper limits are reported at $2\sigma$-level.

\subsection{Full simulation}
\label{sec:totsim}

This procedure, consisting in the simulation both of the background and the source, is crucial to set up our Python code for the analysis.
We simulated an image of $46 \times 46$ pixels, corresponding to $27.6$~arcmin (the pixel size is $36$~arcsec), with a fake source at $x_0 = y_0 = 23$ with $A$ increasing from $0.1$~mJy with step $0.1$~mJy; we assume five cases of the background $N_0$ ($10^{-3}$, $1$, $2$, $5$ and $10$~mJy).
\begin{figure*} 
\centering
{\includegraphics[width=55mm]{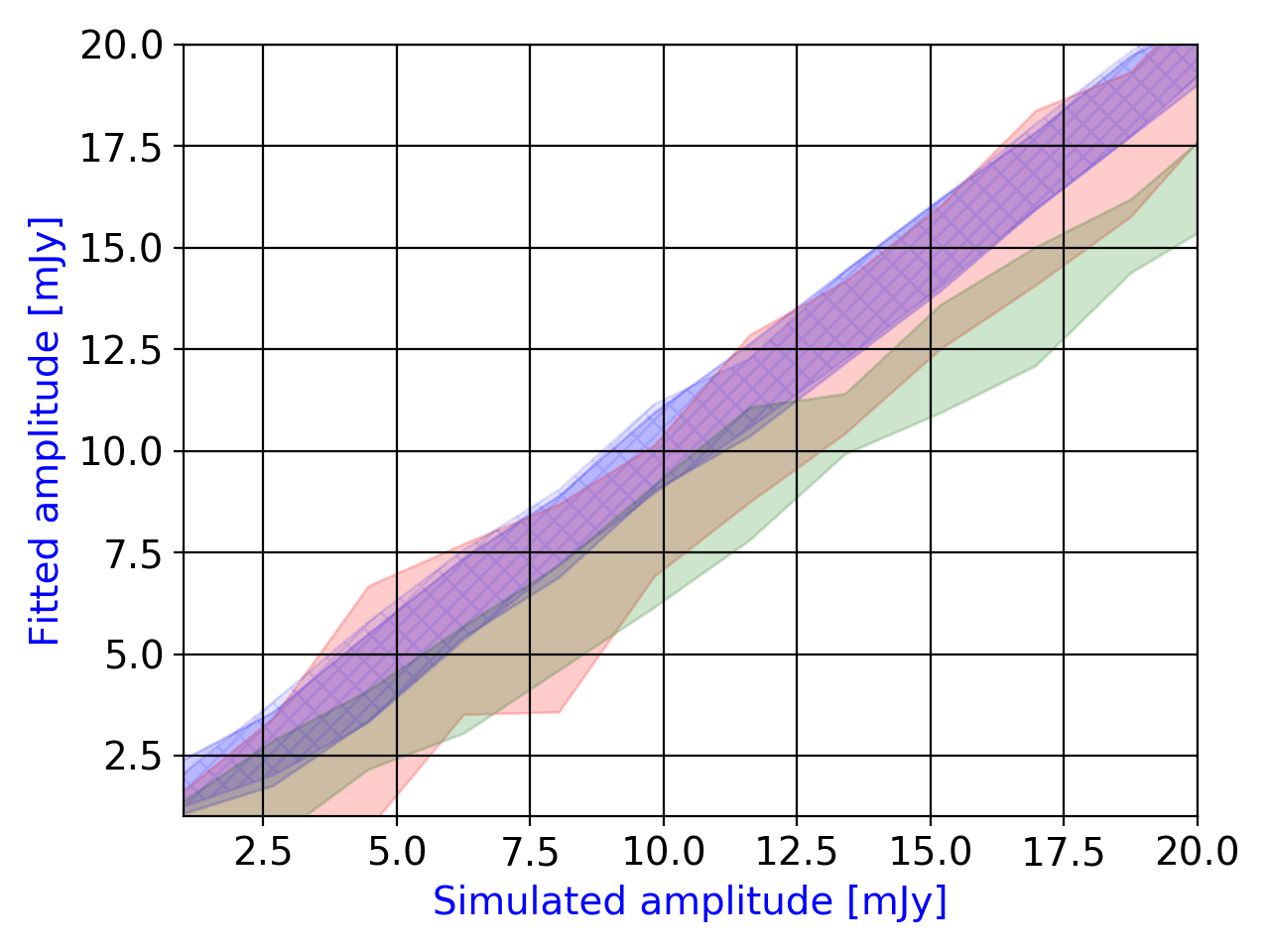}} \quad
{\includegraphics[width=55mm]{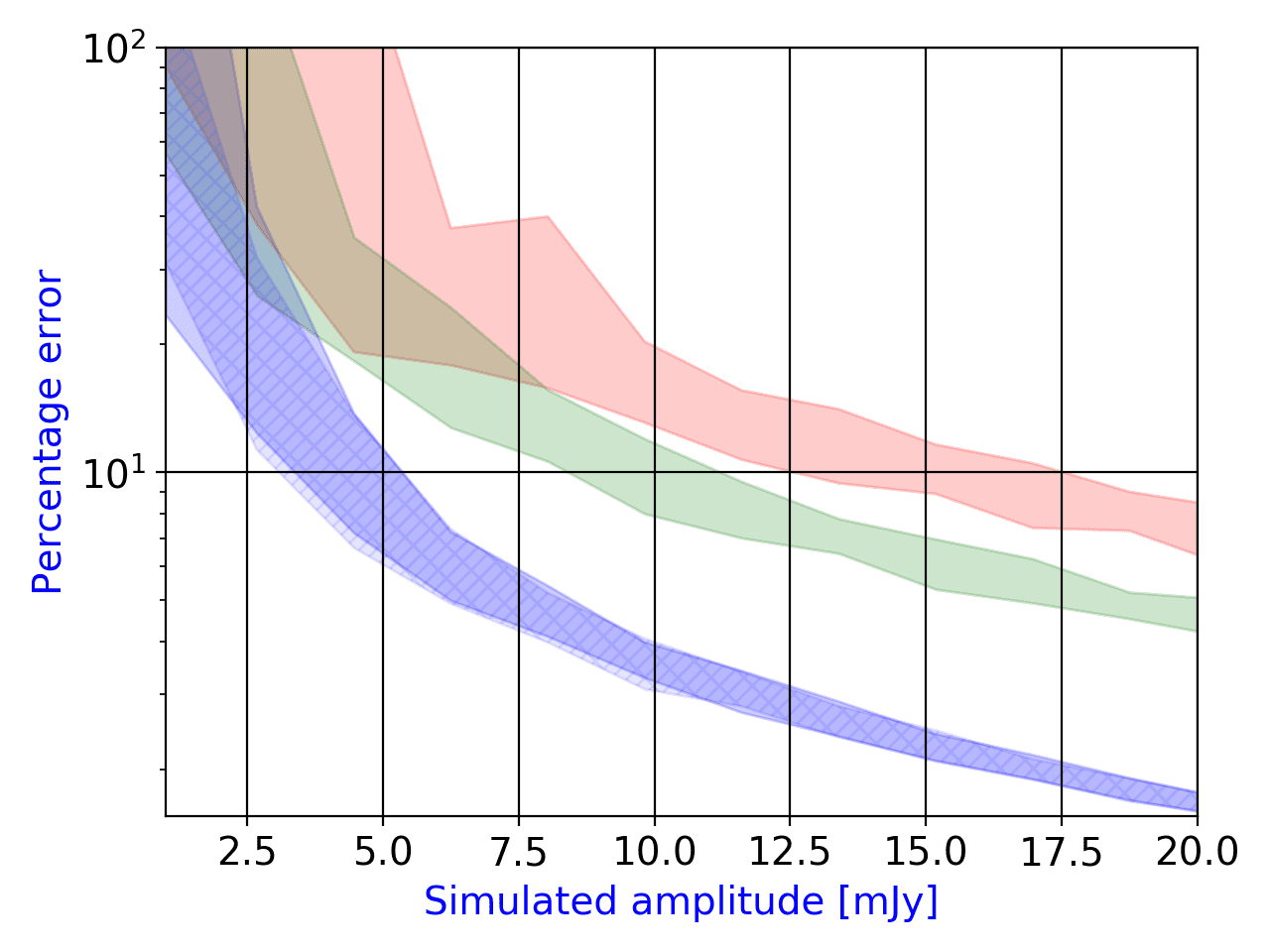}} \\
{\includegraphics[width=55mm]{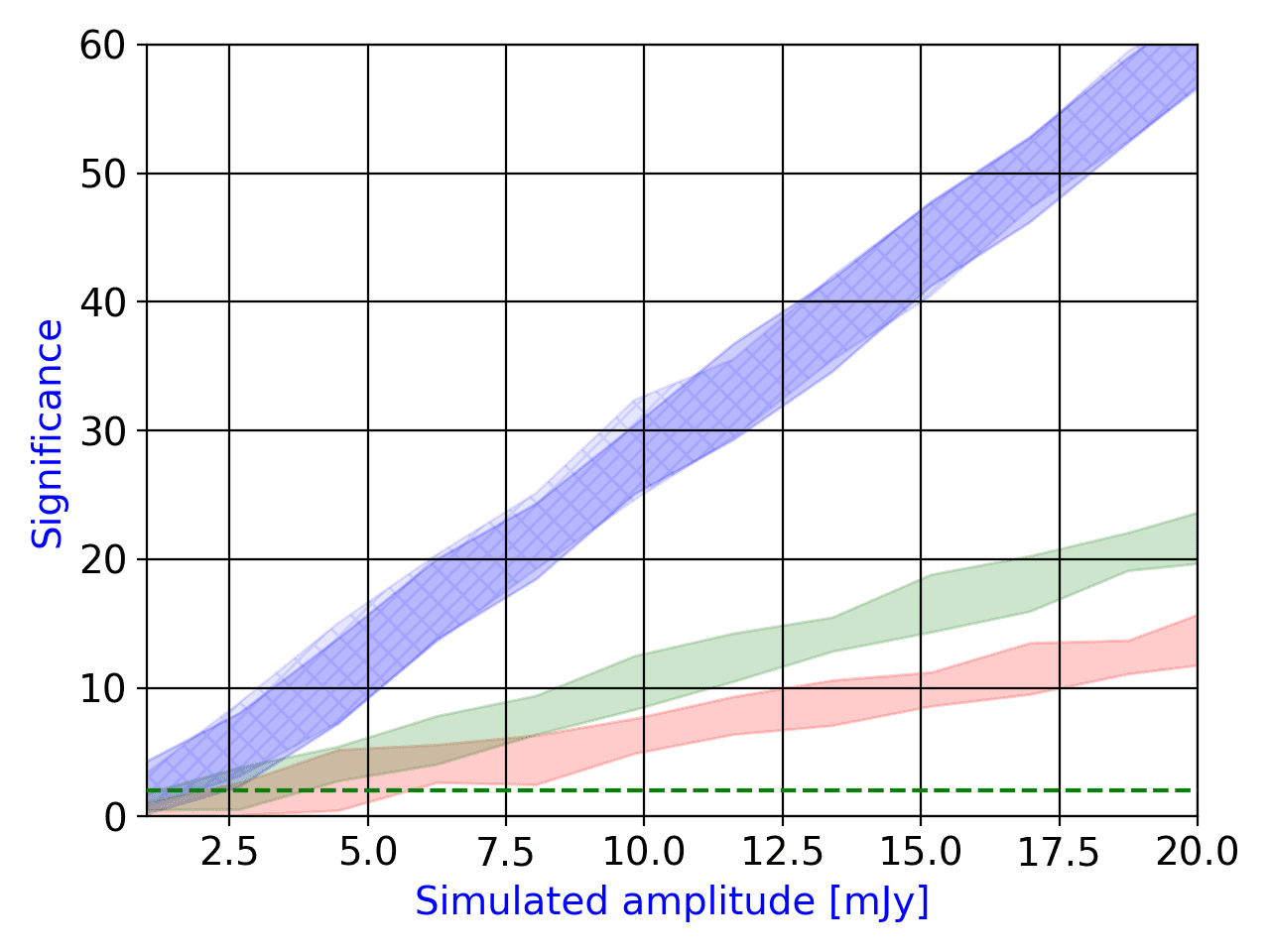}} \quad
{\includegraphics[width=55mm]{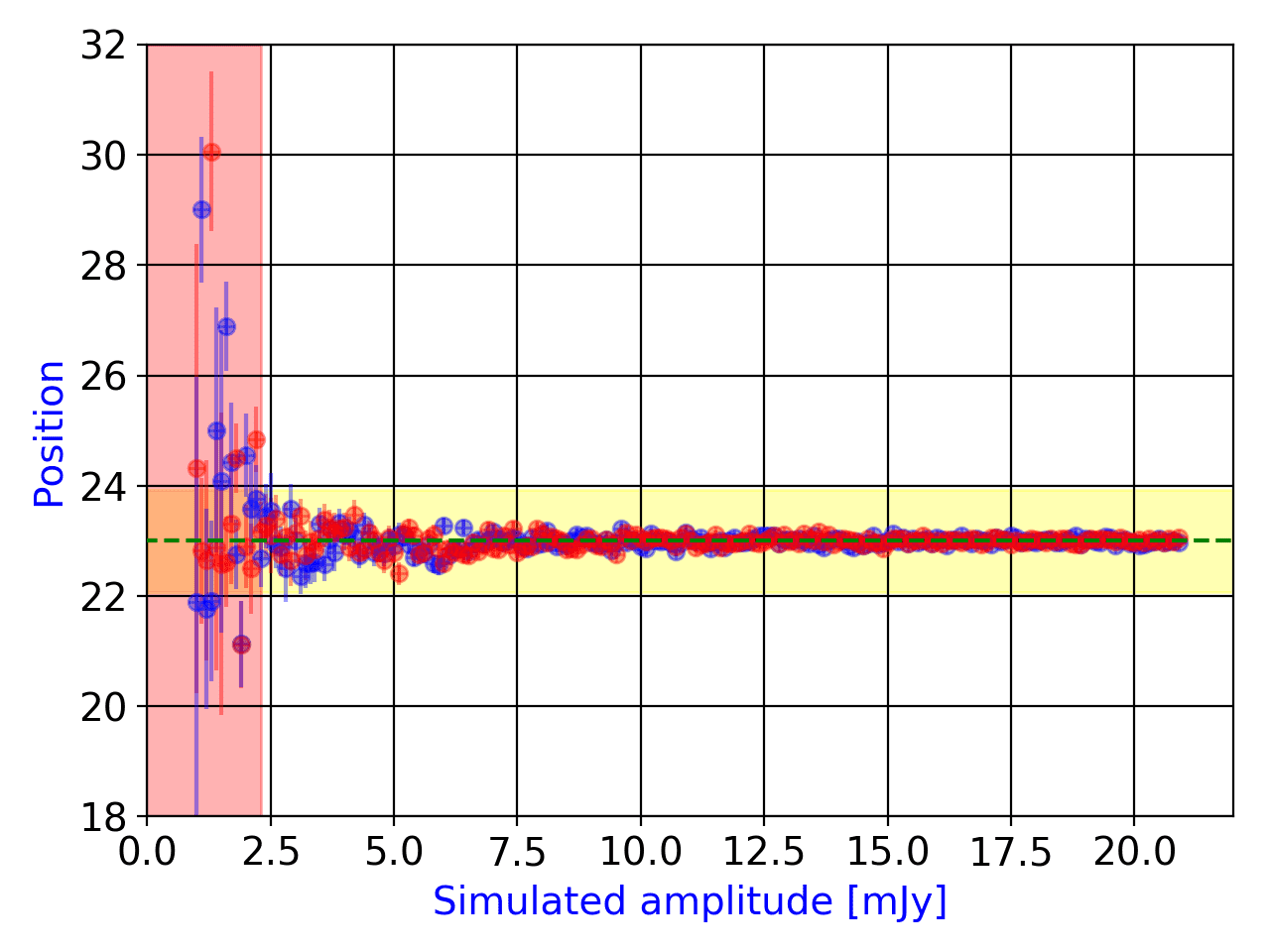}} \\
\caption{Results of full simulation procedure for the case $N_0 = 1$~mJy.
Color shaded area in the panels are bounded by the maximum and the minimum values computed in $20$ bins of flux densities.
Red, green and blue regions are Method A, B and C, respectively; for Method C, uniform region indicates fix position case, and hatched region indicates free position case.
Top left: plot of the fitted amplitude $A_{m}$ as a function of the simulated amplitude $A_0$ (fixed position in blue, free position in red).
Top right: plot of the relative error on the fitted peak as a function of the simulated amplitude $A_0$ (fixed position in blue, free position in red).
Bottom left: plot of the simulated amplitude $A_0$ as a function of the significance $\mathcal{S}$ (fixed position in blue, free position in red); green dotted line indicates the $2\sigma$-level for upper limits.
Bottom right: plot of the fitted position $x_{0}$ (in red) and $y_{0}$ (in blue), for the case of free position, as a function of the simulated amplitude $A_0$; green line indicates the true position of the fake source; yellow area indicates the $4 \%$ region of good positional detection of the source, and red area indicates the excluded regions from the positional criteria.
}
\label{fig:sim_tot_rms1}
\end{figure*}
\begin{figure*} 
\centering
{\includegraphics[width=55mm]{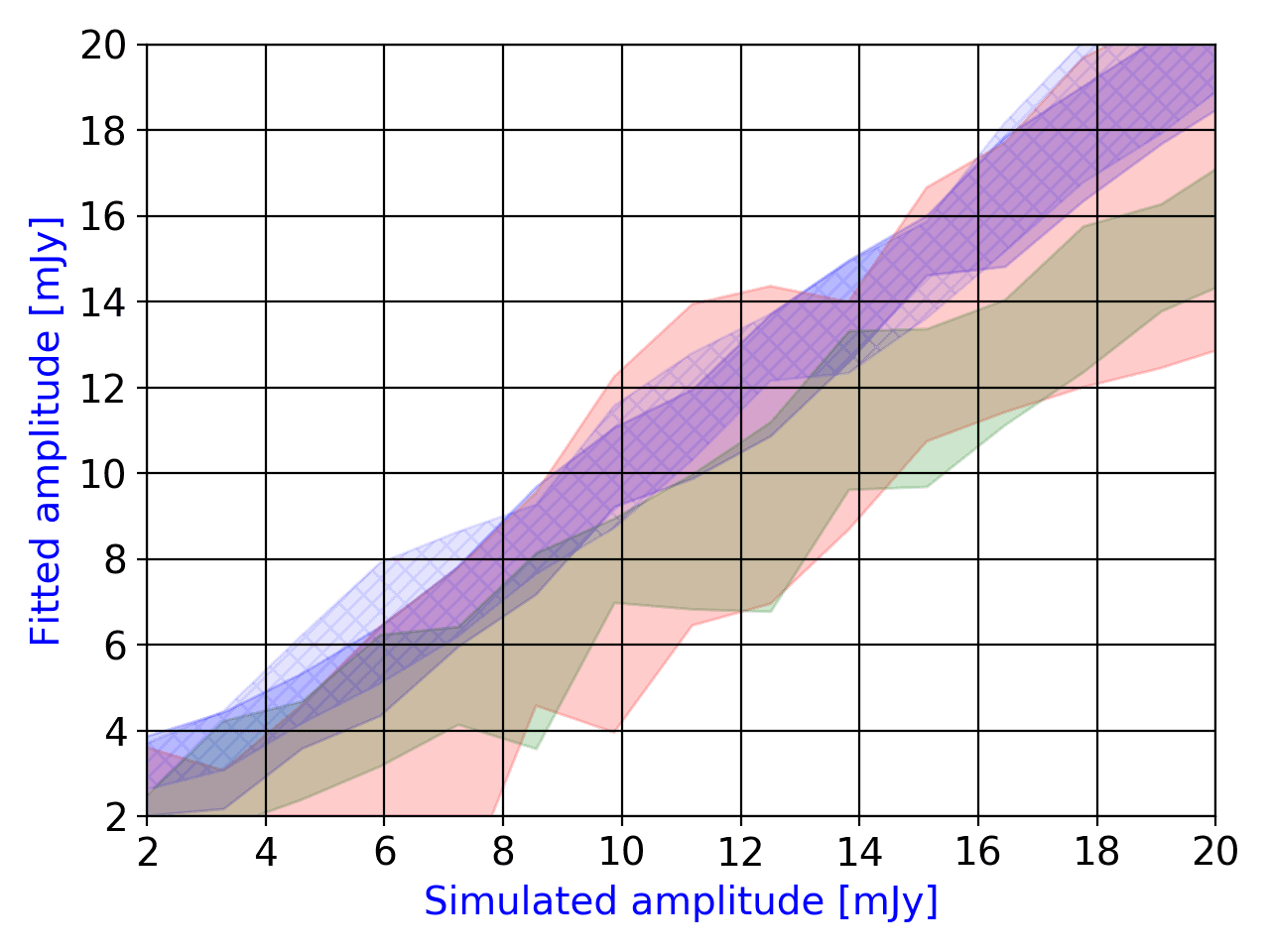}} \quad
{\includegraphics[width=55mm]{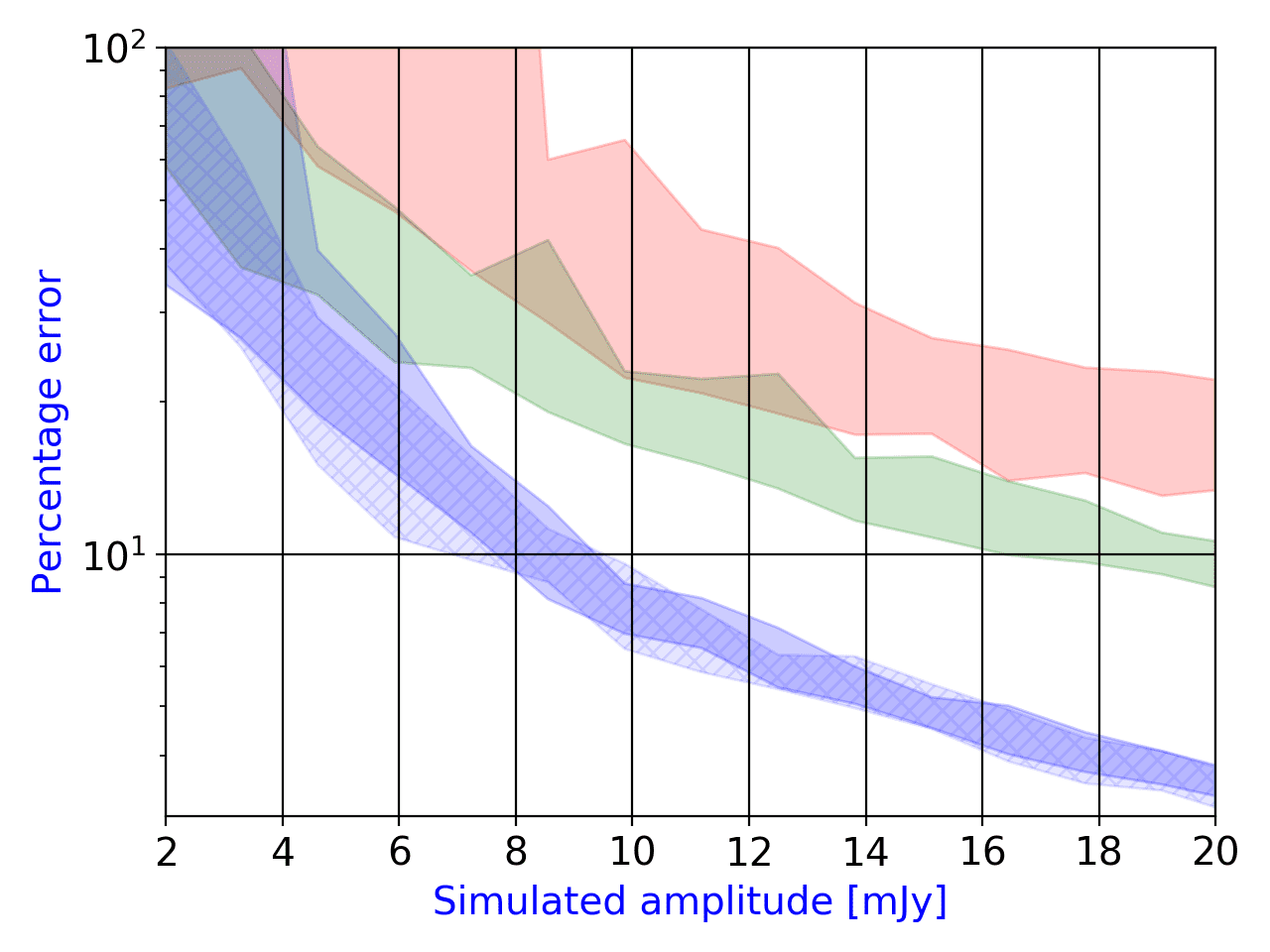}} \\
{\includegraphics[width=55mm]{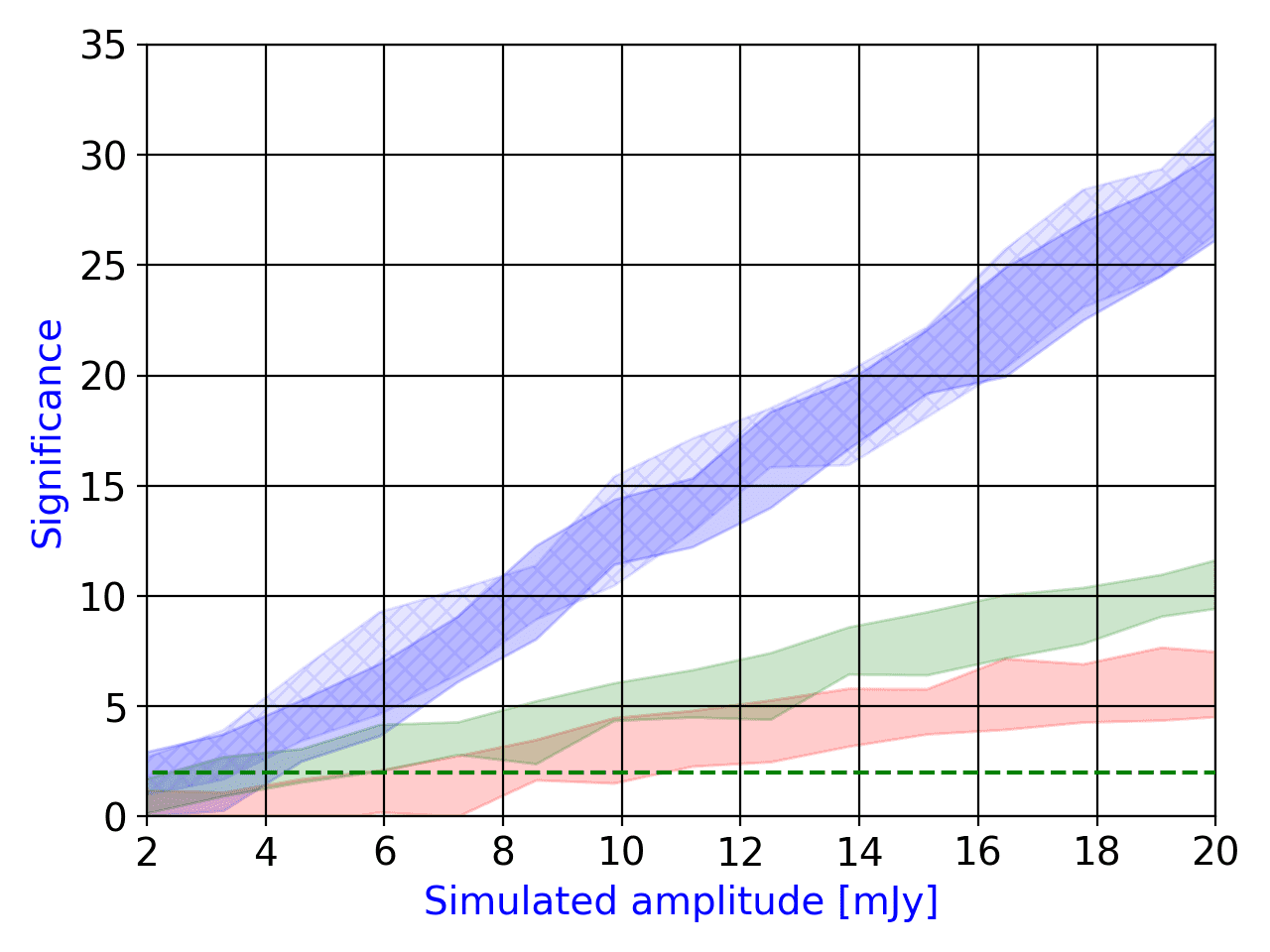}} \quad
{\includegraphics[width=55mm]{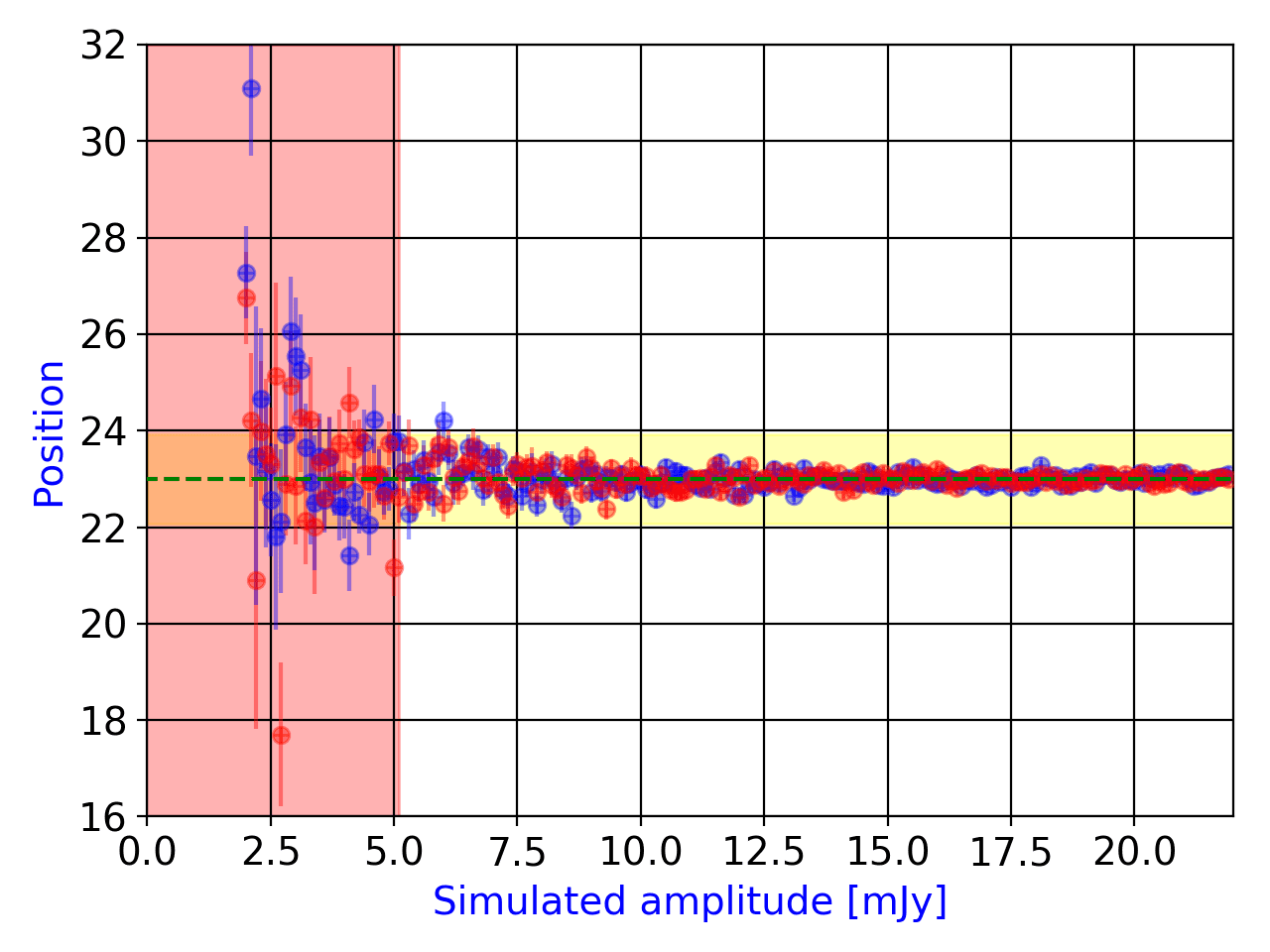}} \\
\caption{
Results of full simulation procedure for the case $N_0 = 2$~mJy.
See the caption of Fig.~\ref{fig:sim_tot_rms1} for a full description of the symbols and plots.
}
\label{fig:sim_tot_rms2}
\end{figure*}
\begin{figure*} 
\centering
{\includegraphics[width=55mm]{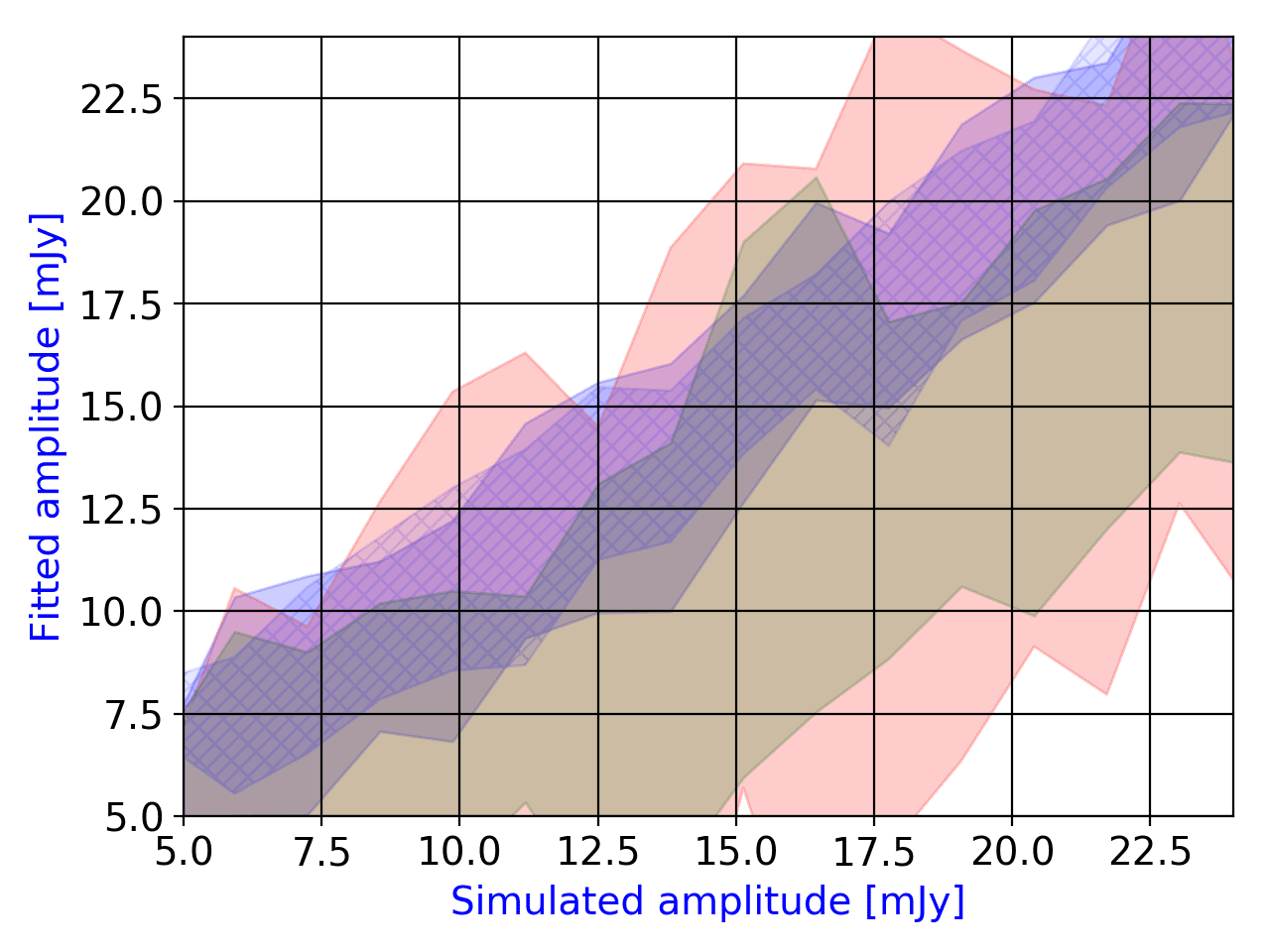}} \quad
{\includegraphics[width=55mm]{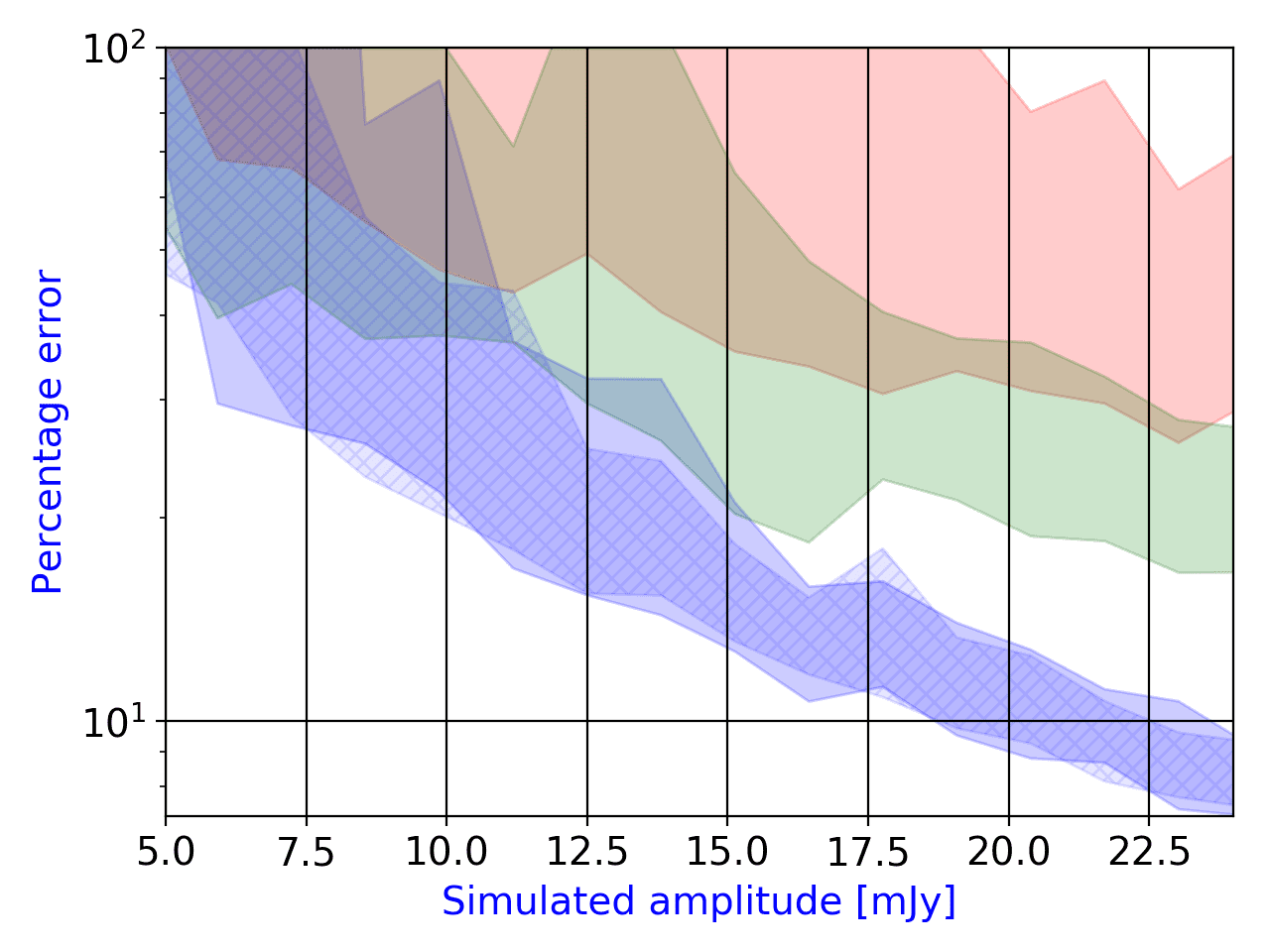}} \\
{\includegraphics[width=55mm]{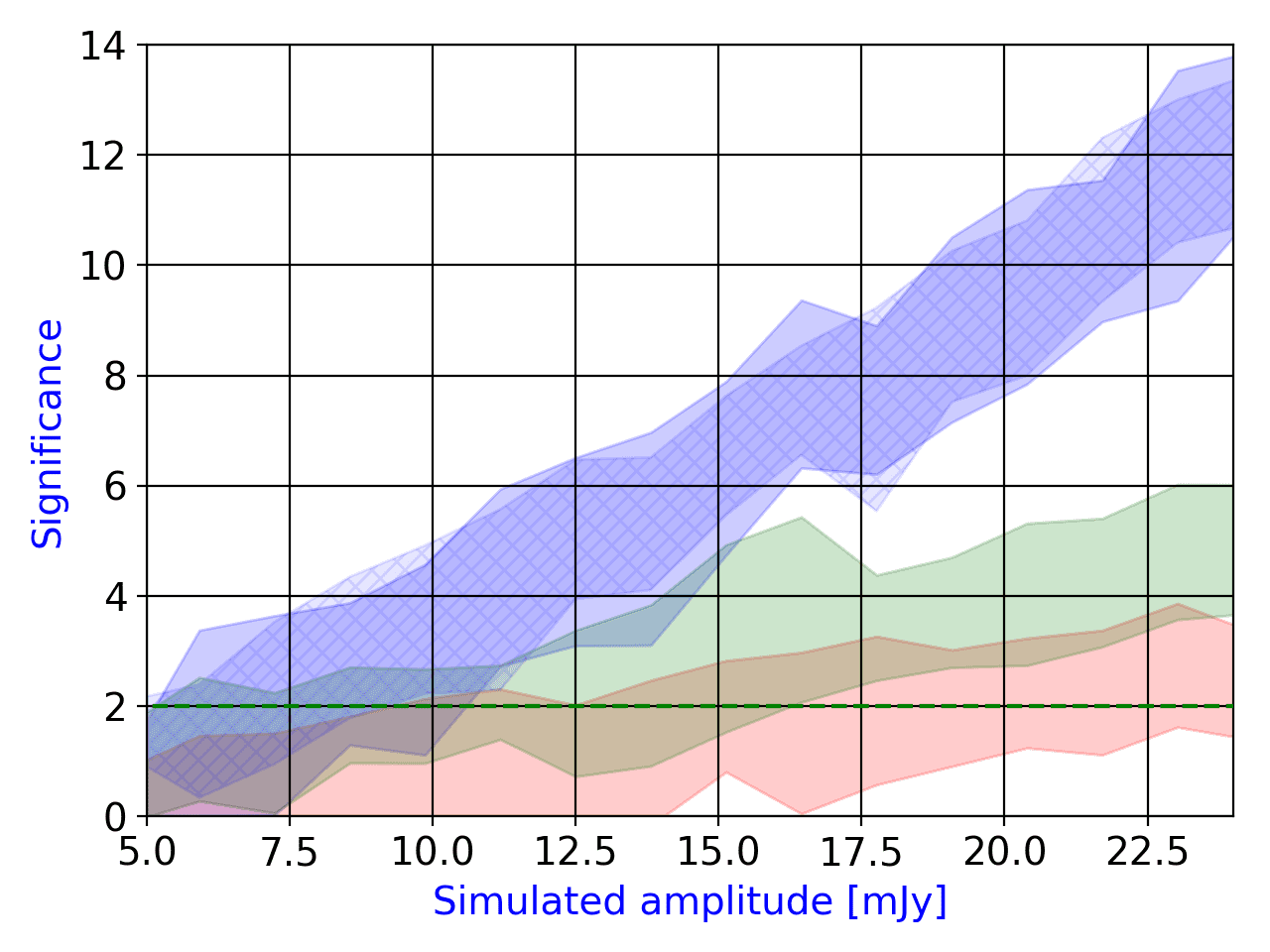}} \quad
{\includegraphics[width=55mm]{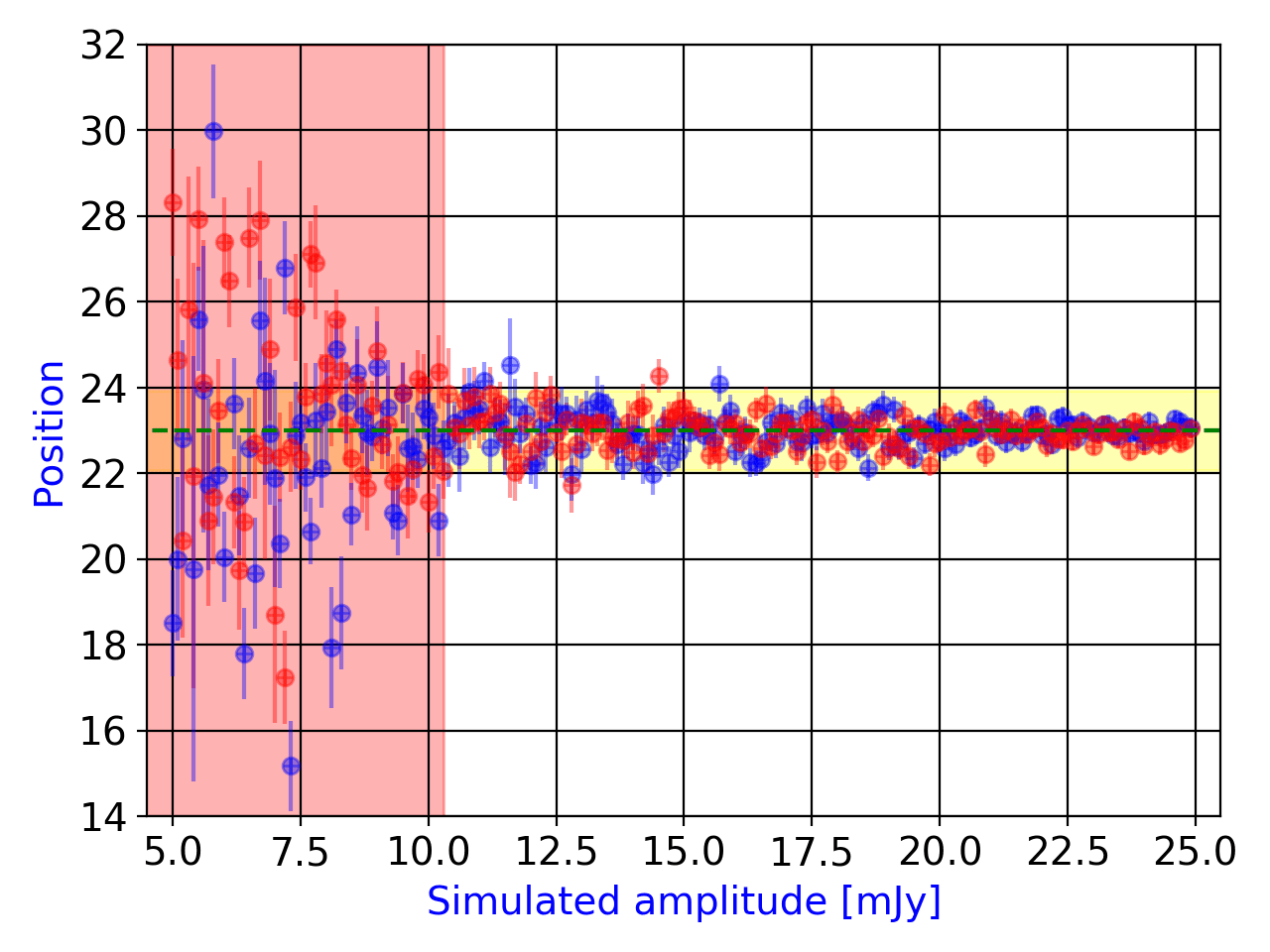}} \\
\caption{
Results of full simulation procedure for the case $N_0 = 5$~mJy.
See the caption of Fig.~\ref{fig:sim_tot_rms1} for a full description of the symbols and plots.
}
\label{fig:sim_tot_rms5}
\end{figure*}
\begin{figure*} 
\centering
{\includegraphics[width=55mm]{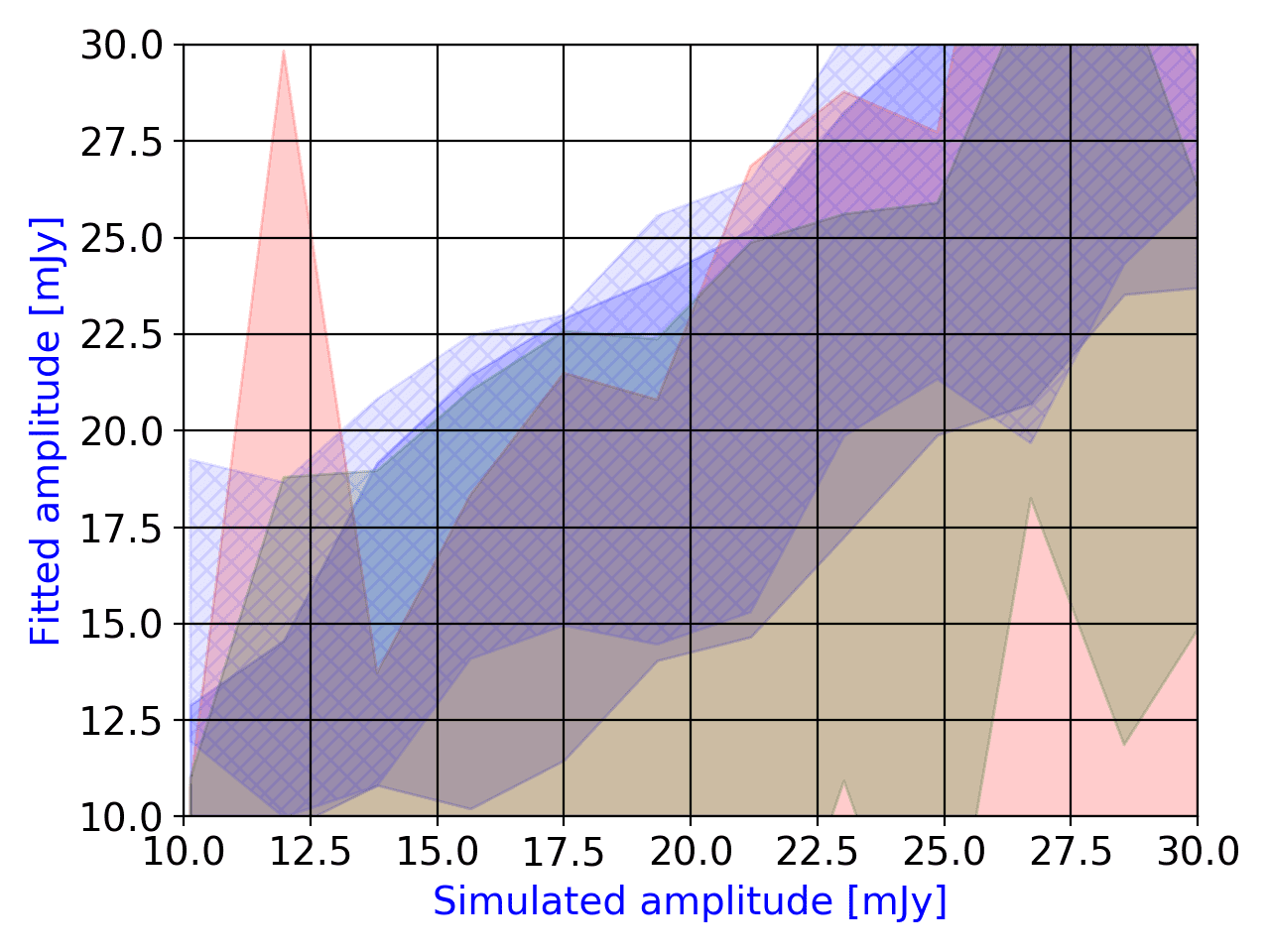}} \quad
{\includegraphics[width=55mm]{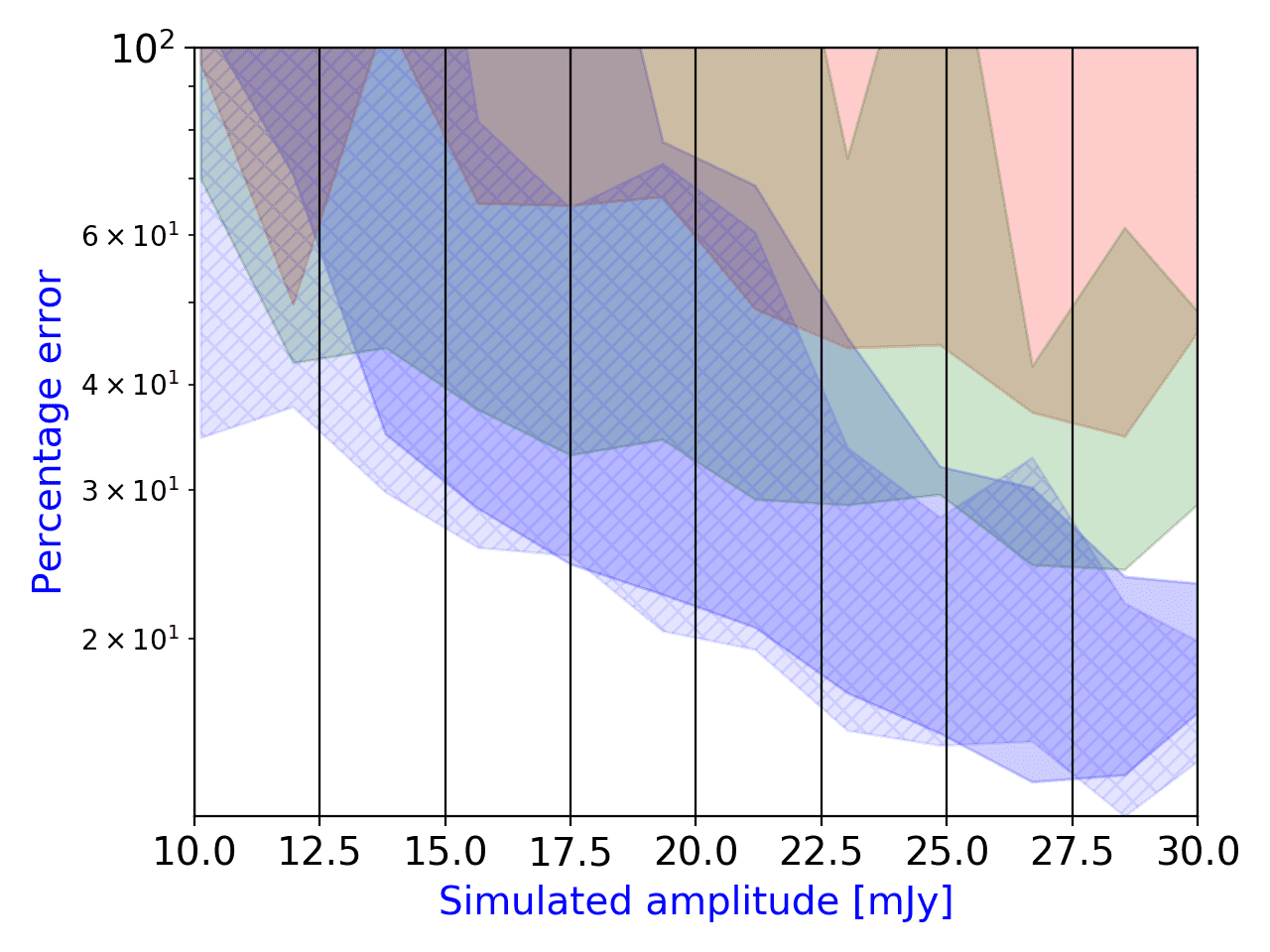}} \\
{\includegraphics[width=55mm]{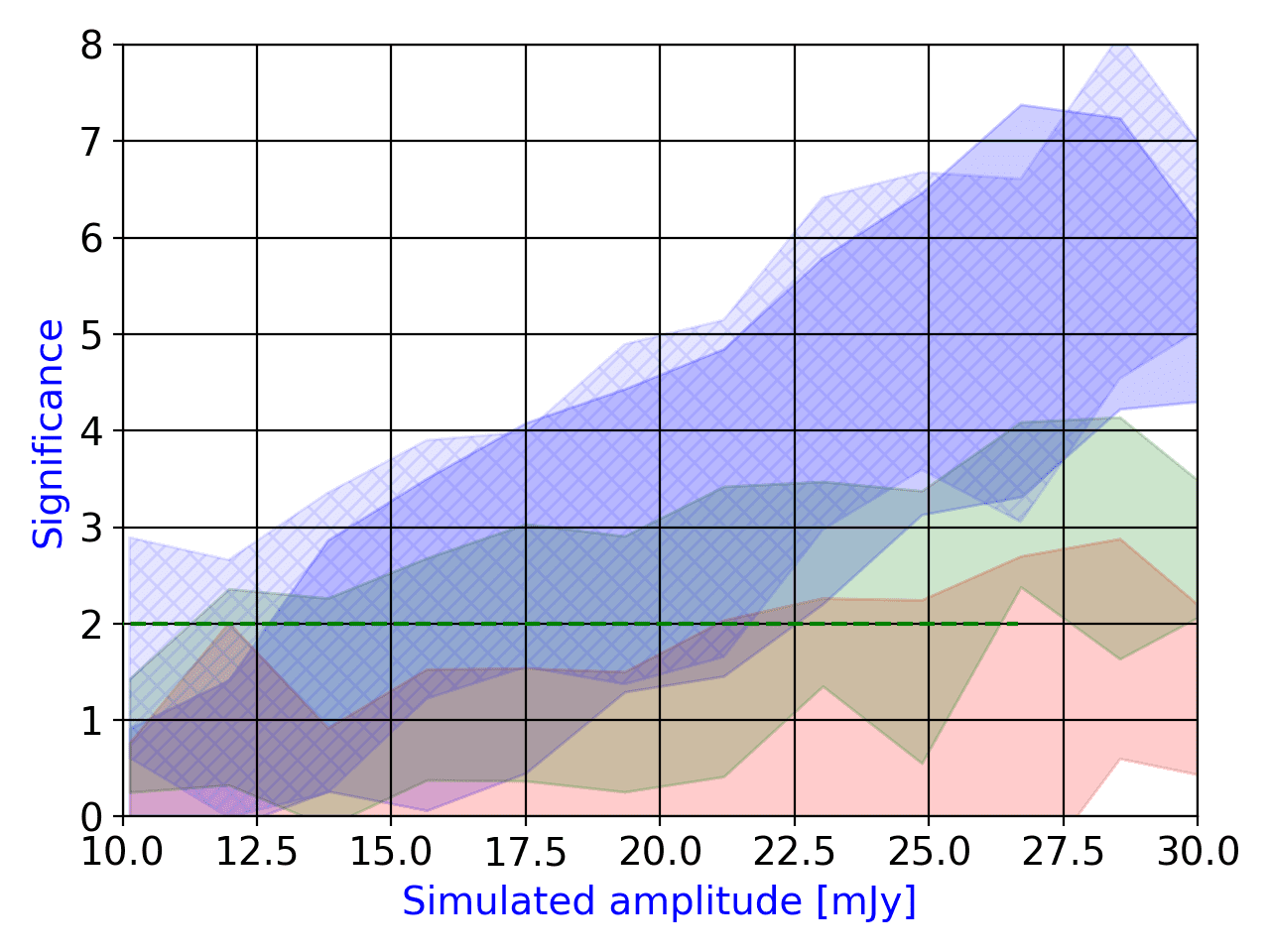}} \quad
{\includegraphics[width=55mm]{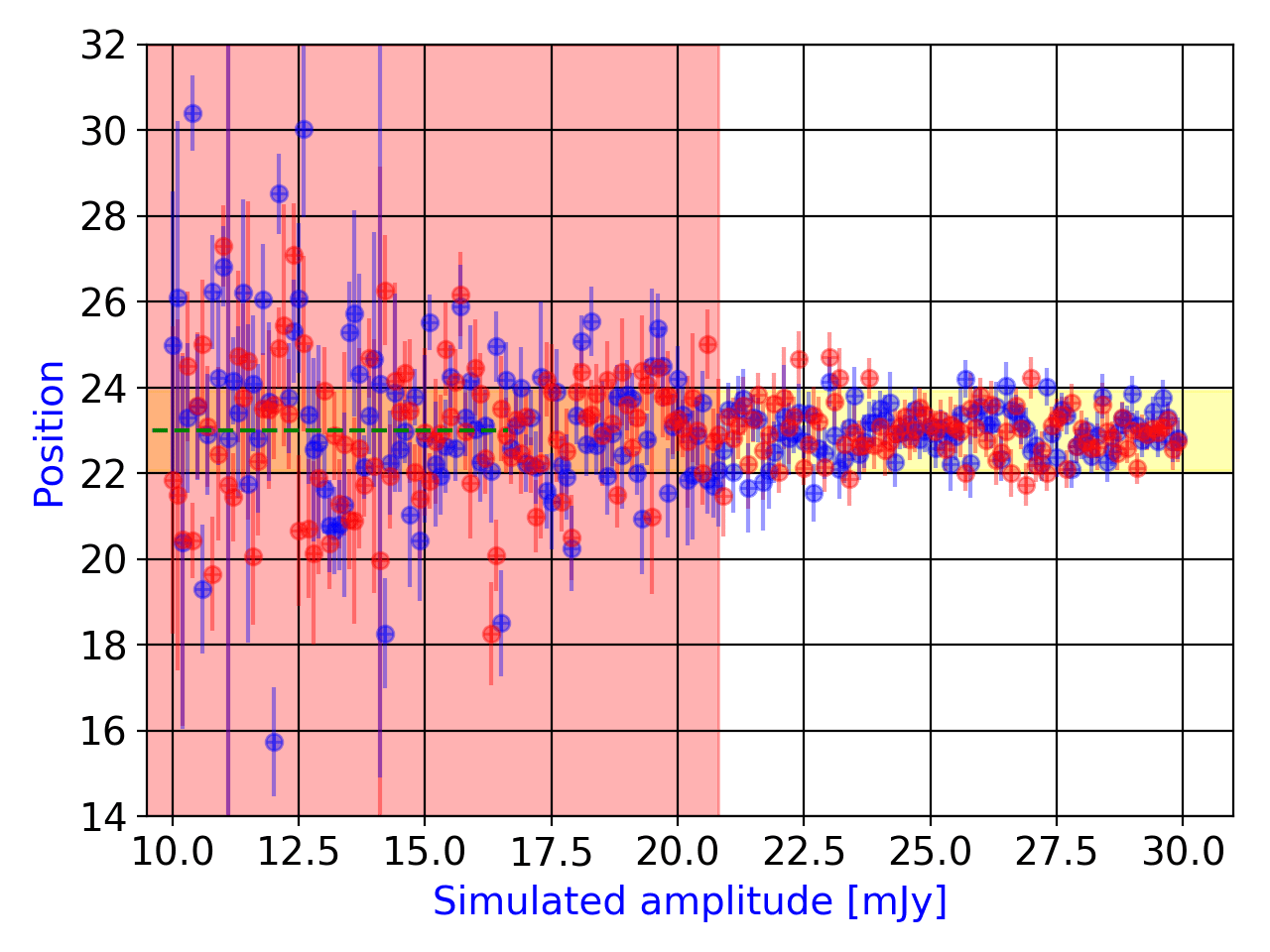}} \\
\caption{
Results of full simulation procedure for the case $N_0 = 10$~mJy.
See the caption of Fig.~\ref{fig:sim_tot_rms1} for a full description of the symbols and plots.
}
\label{fig:sim_tot_rms10}
\end{figure*}

The results shown in Figs.~\ref{fig:sim_tot_rms1}, \ref{fig:sim_tot_rms2}, \ref{fig:sim_tot_rms5}, \ref{fig:sim_tot_rms10}, suggest that the Method C provides an excellent accordance with the original flux density injected in the 2D-Gaussian fake source, implying a good significance $\mathcal{S}$; on the other hand, method A shows high uncertainties and we recommended it only for a rapid and preliminary estimation of the flux density and/or when the instrument beam is poorly known.
According to the best case (Method C), in the fixed position case the $3\sigma$-level detection is reached for flux densities ranging between $\gtrsim 2$~mJy ($N_0 = 1$~mJy, Fig.~\ref{fig:sim_tot_rms1}) and $\gtrsim 15$~mJy ($N_0 = 10$~mJy, Fig.~\ref{fig:sim_tot_rms10}); in the free position case, this detection ranges between $\gtrsim 2.3$~mJy ($N_0 = 1$~mJy, Fig.~\ref{fig:sim_tot_rms1}) and $\gtrsim 20.8$~mJy ($N_0 = 10$~mJy, Fig.~\ref{fig:sim_tot_rms10}); moreover, the ideal case $N_0 = 10^{-3} \sim 0$~mJy corresponds to a source $3\sigma$-level detection at the limiting sensitivity of $\gtrsim 0.1$~mJy.

For Method B, simulations suggest an optimal extraction region (providing the maximum $S/N$ ratio) of $\sim 2$~HPBW; this corresponds to $93.75 \%$ of the Gaussian beam solid angle, and hence for real observations we apply a corrective factor of $1.07$ to the flux densities obtained with this method.

\section{Detection methods applied to real observations}
\label{sec:reale_tutto}

The analysis of these detection methods in the case of real radio observations shows very interesting aspects about the imaging of faint sources and the upper limit estimation in case of undetected targets.
Fig.~\ref{fig:im_real} (bottom) shows the image of the detected source GRS\,1915+105; our detection methods produce similar values of flux densities (Table~\ref{tab:quickand2}).
In particular, Method C is able to detect a single source in the field also in free position case, despite the region being characterized by variable background and strong RFI.
The analysis on this source (Table~\ref{tab:quickand2}) shows that Method C provides an uncertainty on the flux density measurement of $\sim 5 \%$ ($\sim 7 \%$ for Method B, and $\sim 9 \%$ for Method A), suggesting that this method is the most accurate and robust for flux densities measurements.

The image of GRB\,190114C in Fig.~\ref{fig:im_real} is the prototype of a field characterized by an undetected source; we discuss this part in the following Section.
Table~\ref{tab:quickand2} shows in detail the results of our Python code for real observations.

\subsection{Real images of non-detected sources}
\label{sec:realmaps}

In single-dish mode, often faint radio sources could be non-detectable.
In this context we wonder "What is the minimum significance level needed for source detection?".
To answer this question, we need to inject fake sources in the real radio image (Eq.~\ref{eq:gauss2d}) in order to understand when the target becomes distinguishable from the background.
These fake sources are located in the position detected by other facilities (from radio to high-energy frequencies); we simulated increasing values of $A$ from $0.1$~mJy to the maximum value of flux density in each image, with step $0.1$~mJy.
A detection of the fake injected source at $2\sigma$-level sets the precise upper limit in the real image at the actual source position. 

\begin{figure*} 
\centering
{\includegraphics[width=55mm]{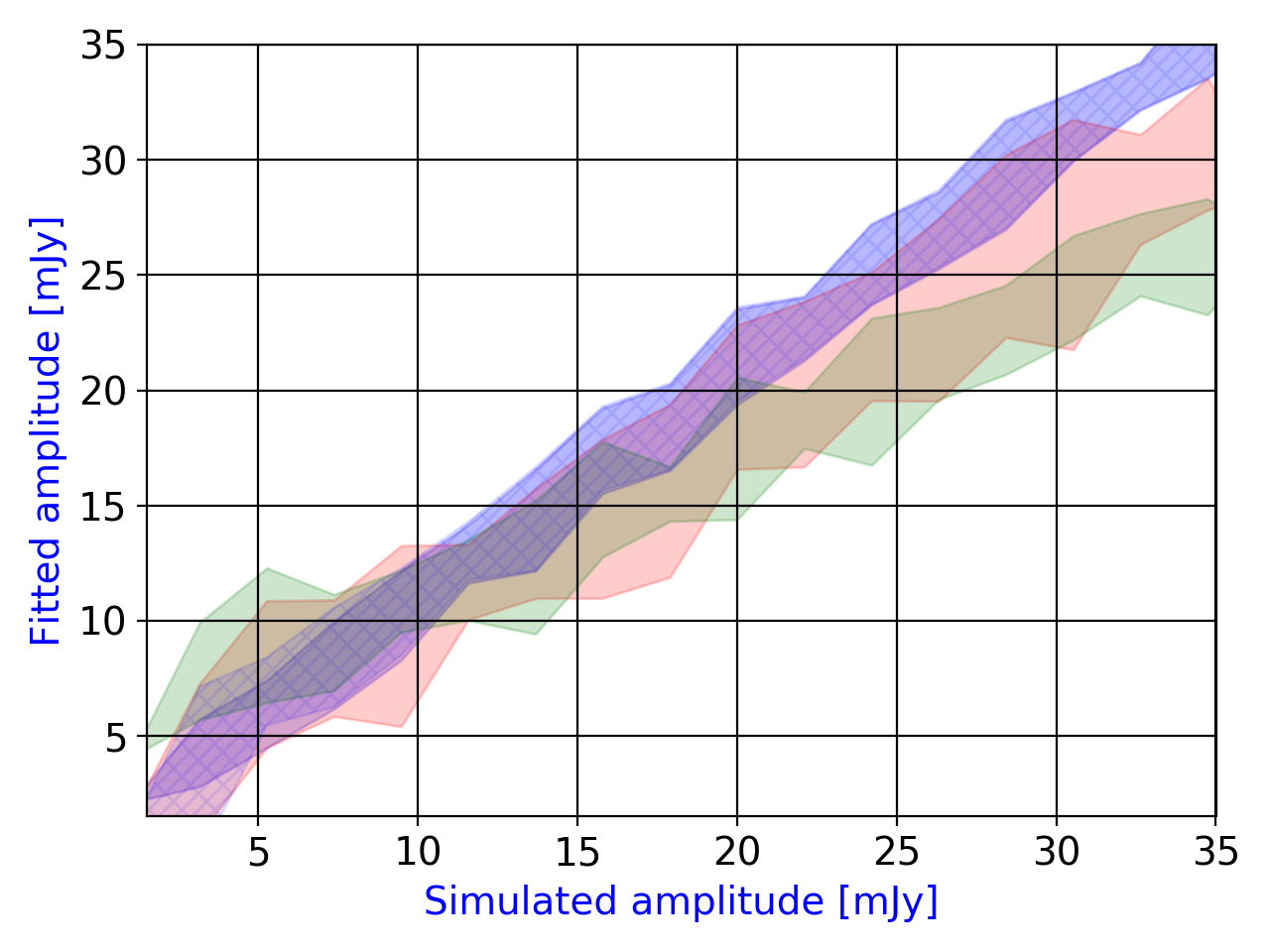}} \quad
{\includegraphics[width=55mm]{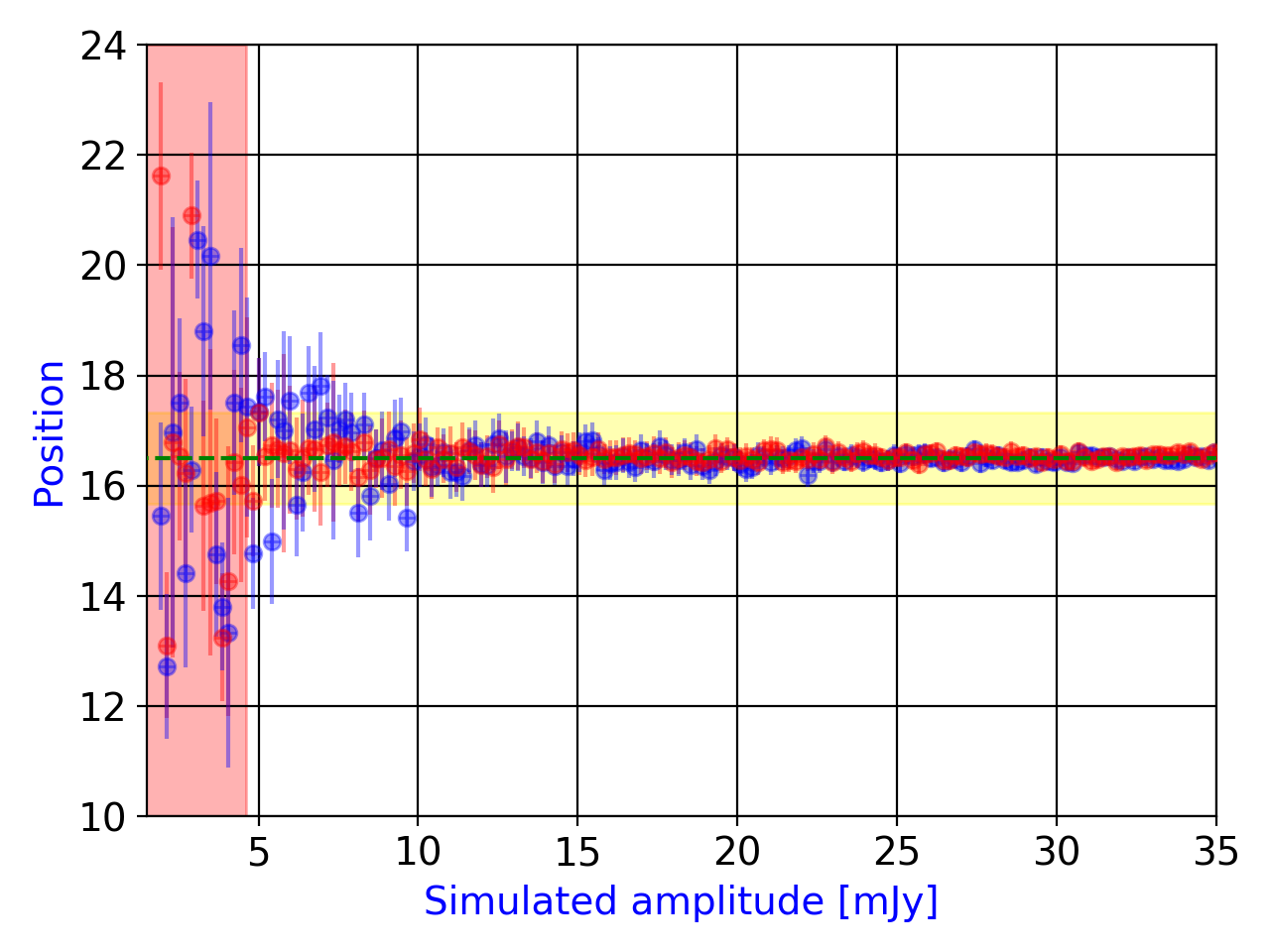}} \\
{\includegraphics[width=55mm]{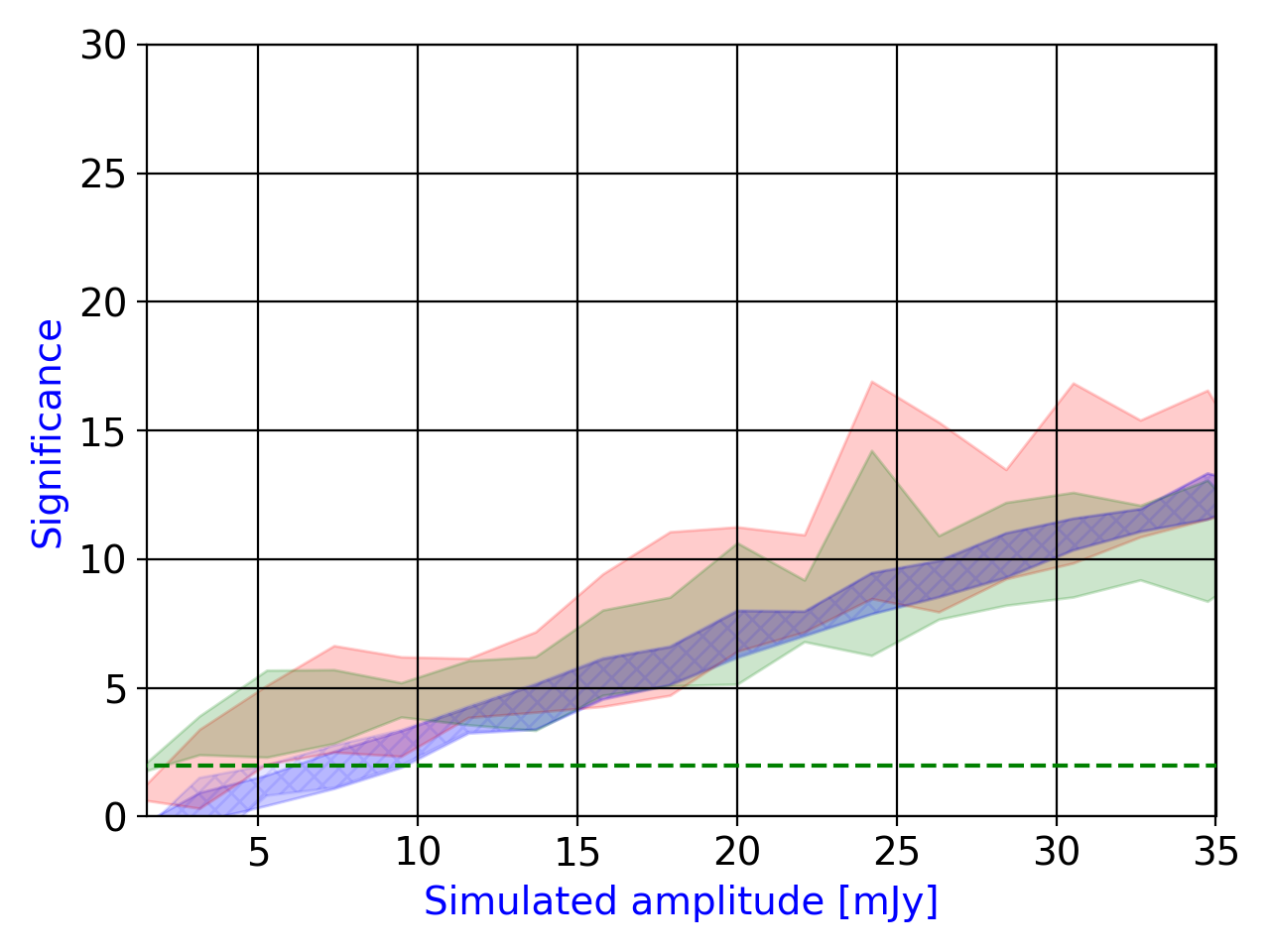}} \quad
{\includegraphics[width=55mm]{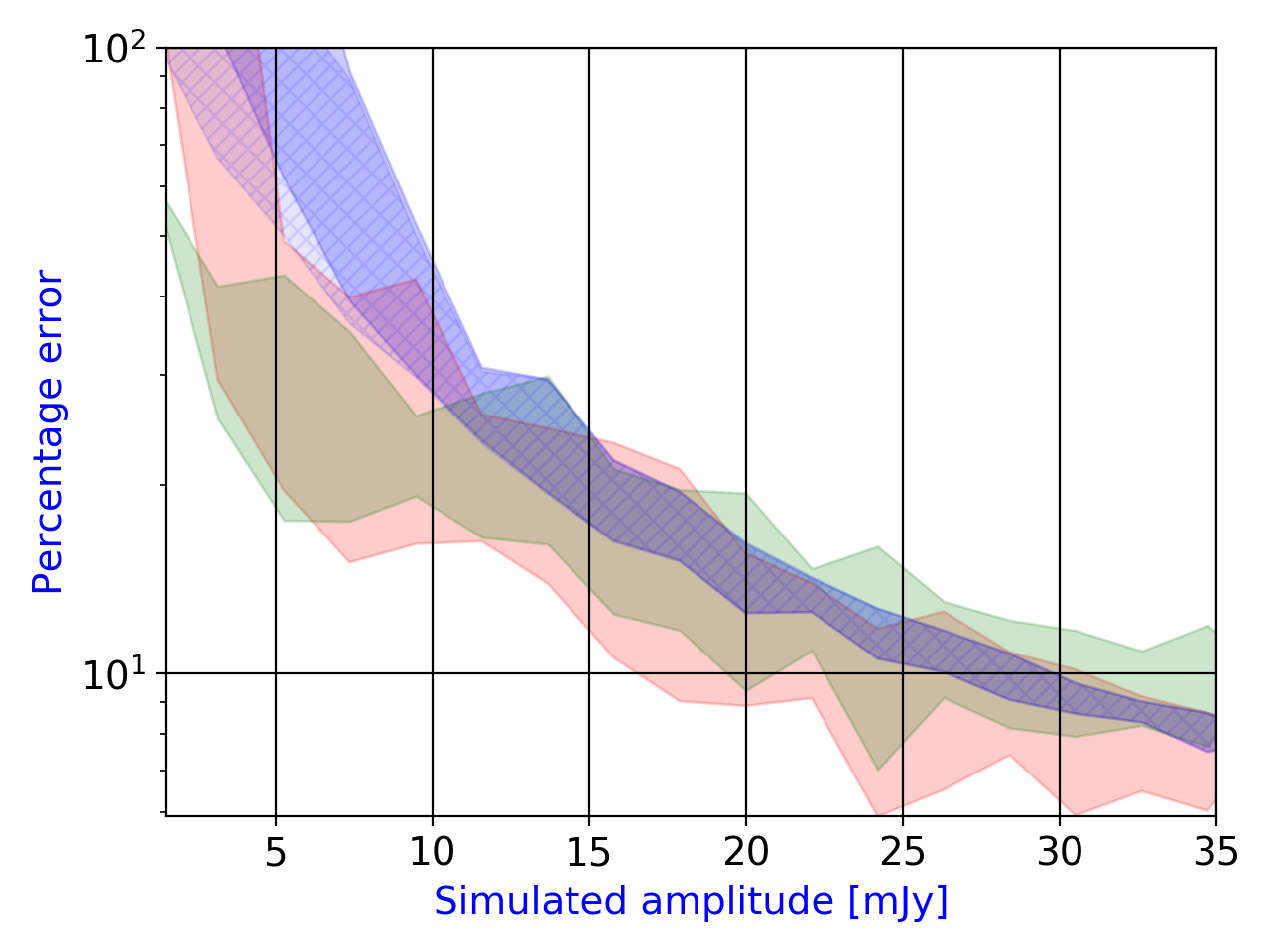}} \\
\caption{Overview of fitting procedure for a fake source (suited by a 2D-Gaussian) about SRT observation of GRB\,181201A at 2018 December 11.
See the caption of Fig.~\ref{fig:sim_tot_rms1} for a full description of the symbols and plots.}
\label{fig:20181211}
\end{figure*}

The first epoch at 2018 December 11 shows $RMS_{min} \sim 1.9$~mJy (Table~\ref{tab:minrms}), resulting in a standard upper limit estimation of $3.8$~mJy ($2\sigma$-level).
The injection of fake source in this field shows that the $2\sigma$-level upper limit of the source is $\sim 3.9$~mJy for Method A, $\sim 2.2$~mJy for Method B and $\sim 7.4$~mJy for Method C (Fig.~\ref{fig:20181211}).
The additional criterion for the free position case (with a fake source) of Method C shows a detection at $\sim 4.6$~mJy where, for the Method C, the significance $\mathcal{S}$ is $\sim 5$ ($\sim 5$ for Method A, and $\sim 7$ for method B).
This is the only case where Method C is characterized by $2\sigma$-level upper limit higher than Method A and B estimations: this high value originates probably by an inaccurate baseline subtraction.

\begin{figure*} 
\centering
{\includegraphics[width=55mm]{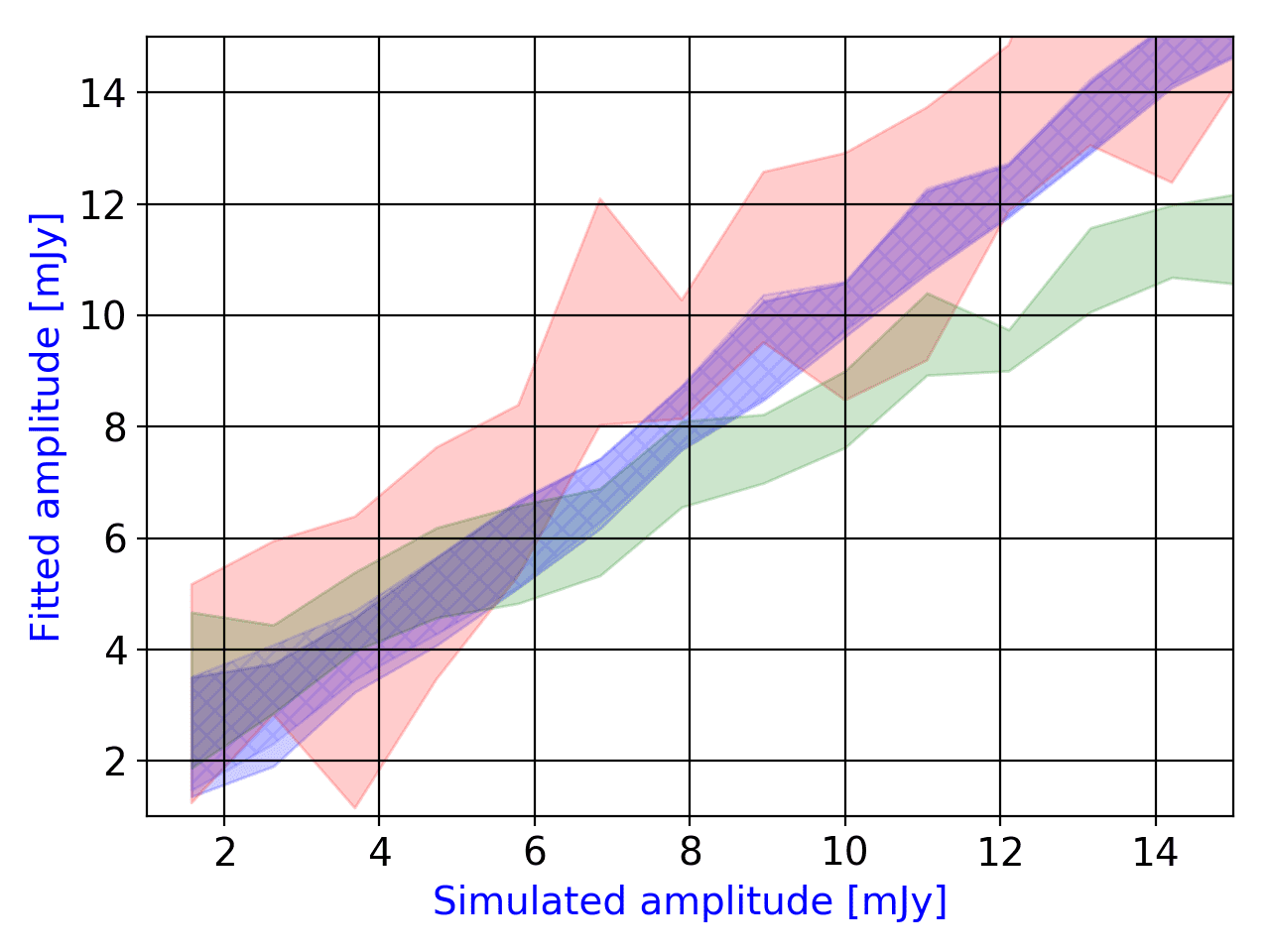}} \quad
{\includegraphics[width=55mm]{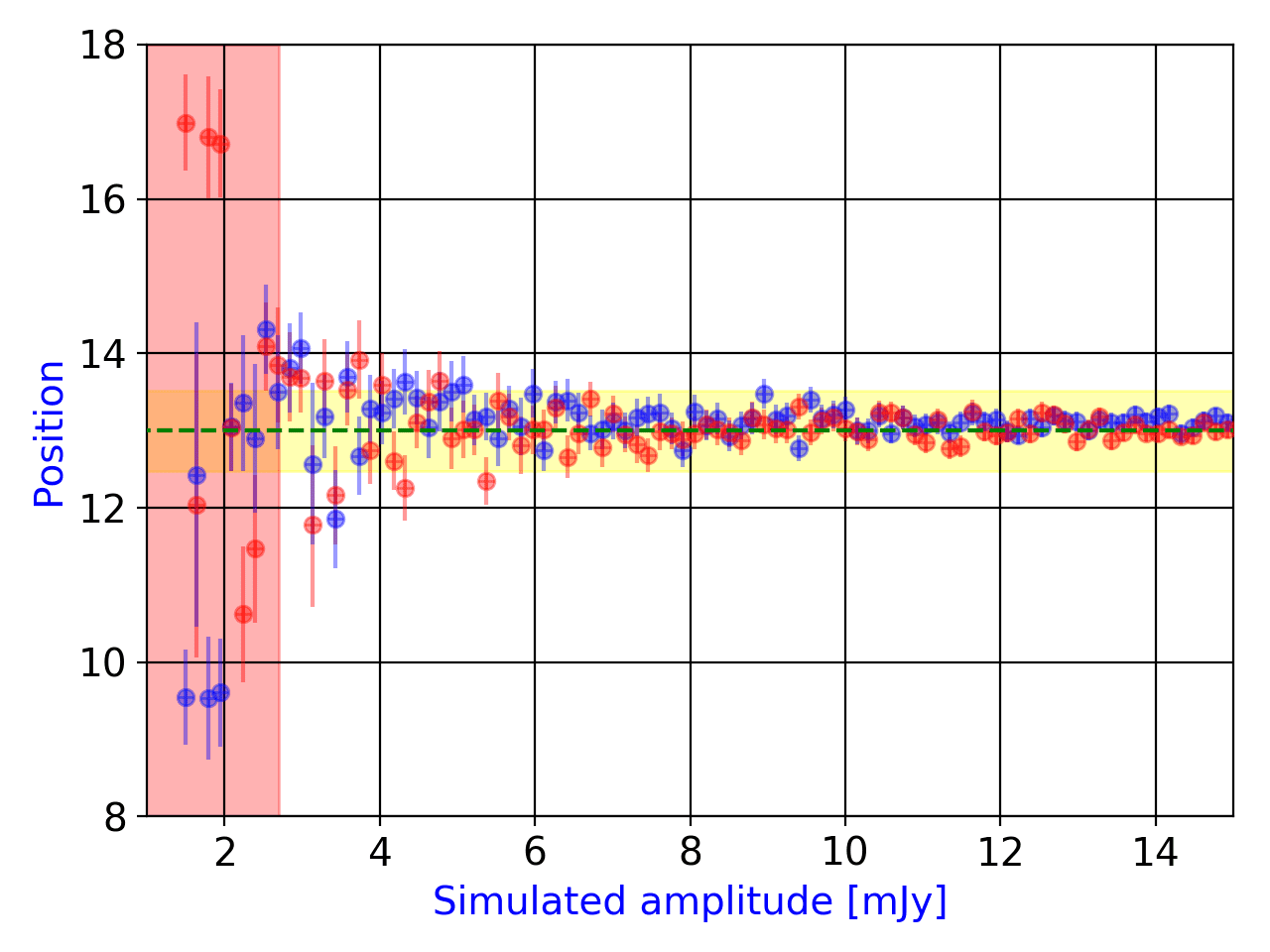}} \\
{\includegraphics[width=55mm]{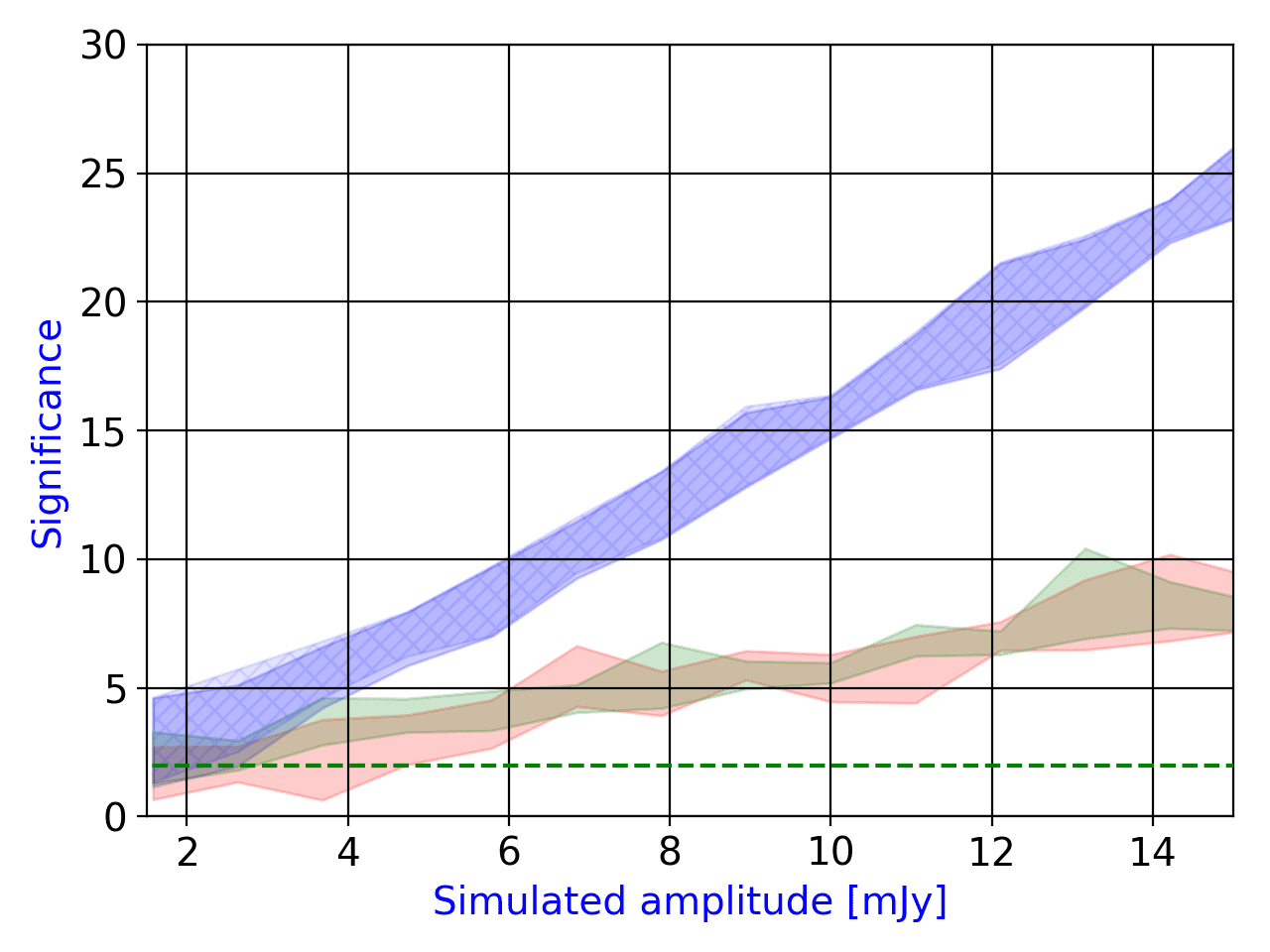}} \quad
{\includegraphics[width=55mm]{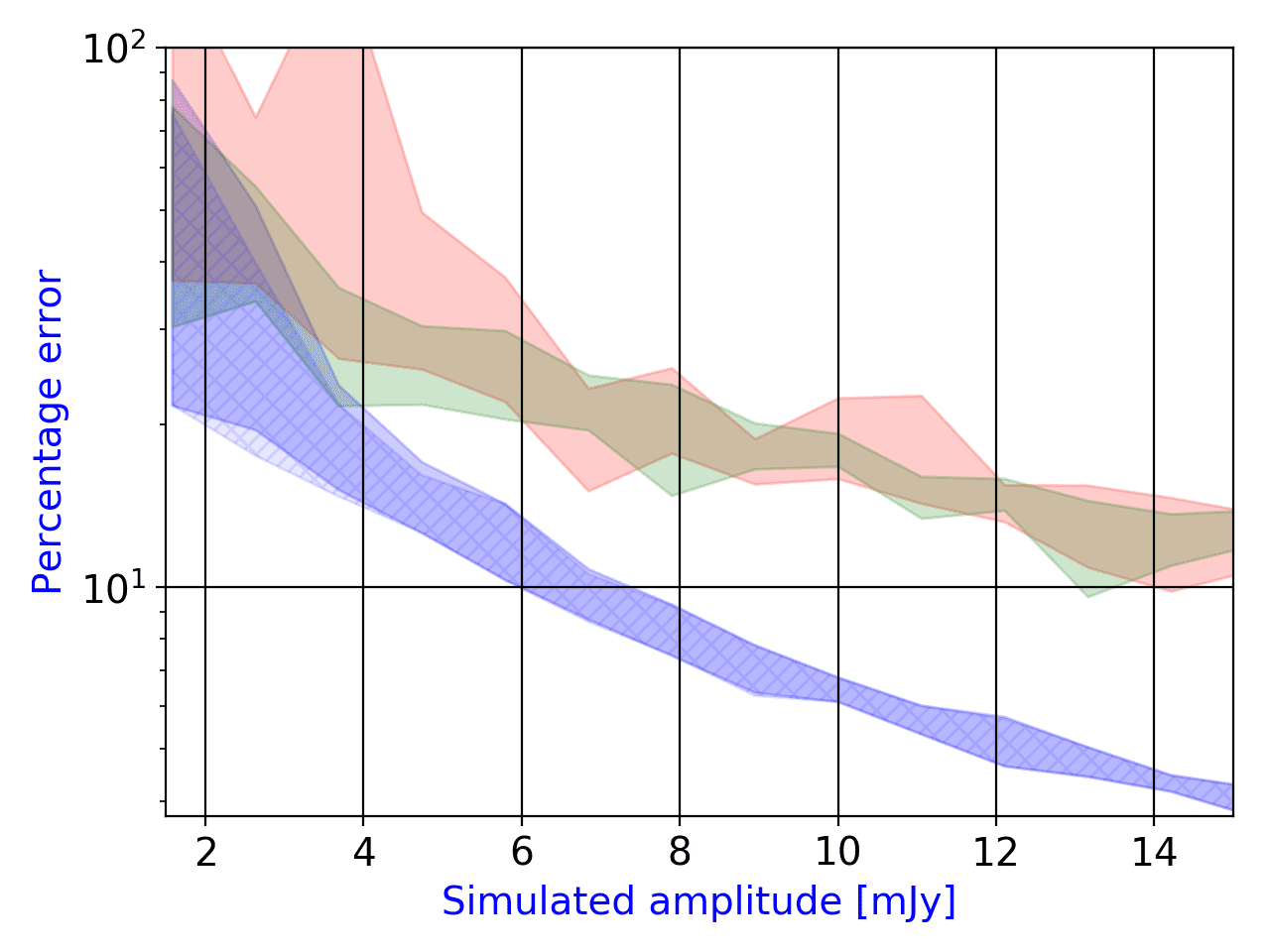}} \\
\caption{Overview of fitting procedure for a fake source (suited by a 2D-Gaussian) about SRT observation of GRB\,190114C at 2019 January 17.
See the caption of Fig.~\ref{fig:sim_tot_rms1} for a full description of the symbols and plots.}
\label{fig:20190117}
\end{figure*}

The second epoch at 2019 January 17 shows a very low $RMS_{min}$ ($\sim 1.5$~mJy, Table~\ref{tab:minrms}), caused probably by the optimal weather conditions at that epoch, resulting in an high image quality.
As we can see in Fig.~\ref{fig:20190117}, we confirm the results for the full simulation procedure: the most accurate method is the method C, and the worst method is the method A (Fig.~\ref{fig:20190117}).
The injection of fake source in this field suggests a $2\sigma$-level upper limit at $\sim 3.4$~mJy for Method A, $\sim 2.5$~mJy for Method B and $\sim 1.8$~mJy for Method C (Fig.~\ref{fig:20190117}); the additional criterion for the free position case for Method C shows a detection at $\sim 2.7$~mJy, where the significance $\mathcal{S}$ is $\sim 4$ ($\sim 2.5$ for both Method A and B; Fig.~\ref{fig:20190117}).

\begin{figure*} 
\centering
{\includegraphics[width=55mm]{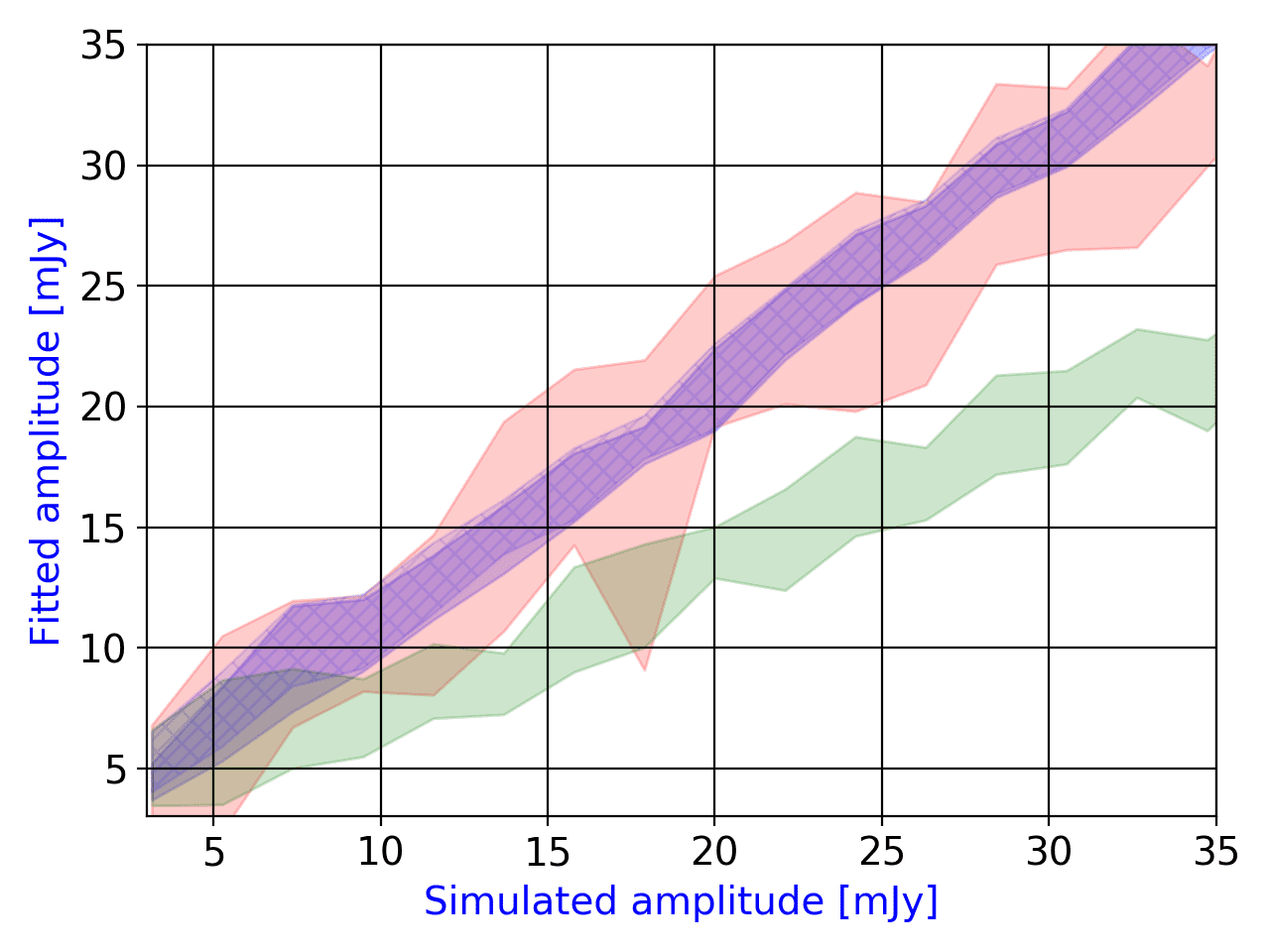}} \quad
{\includegraphics[width=55mm]{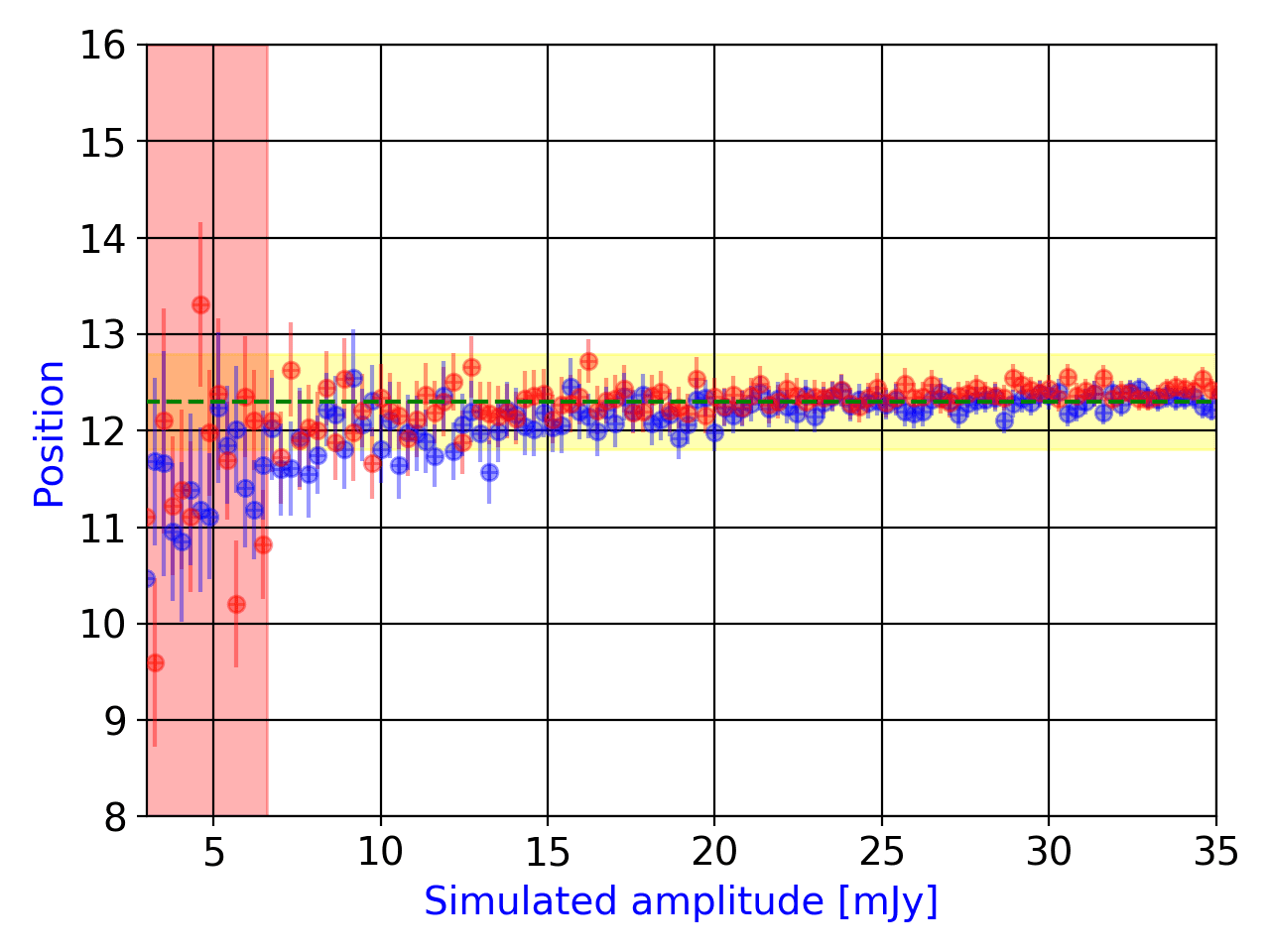}} \\
{\includegraphics[width=55mm]{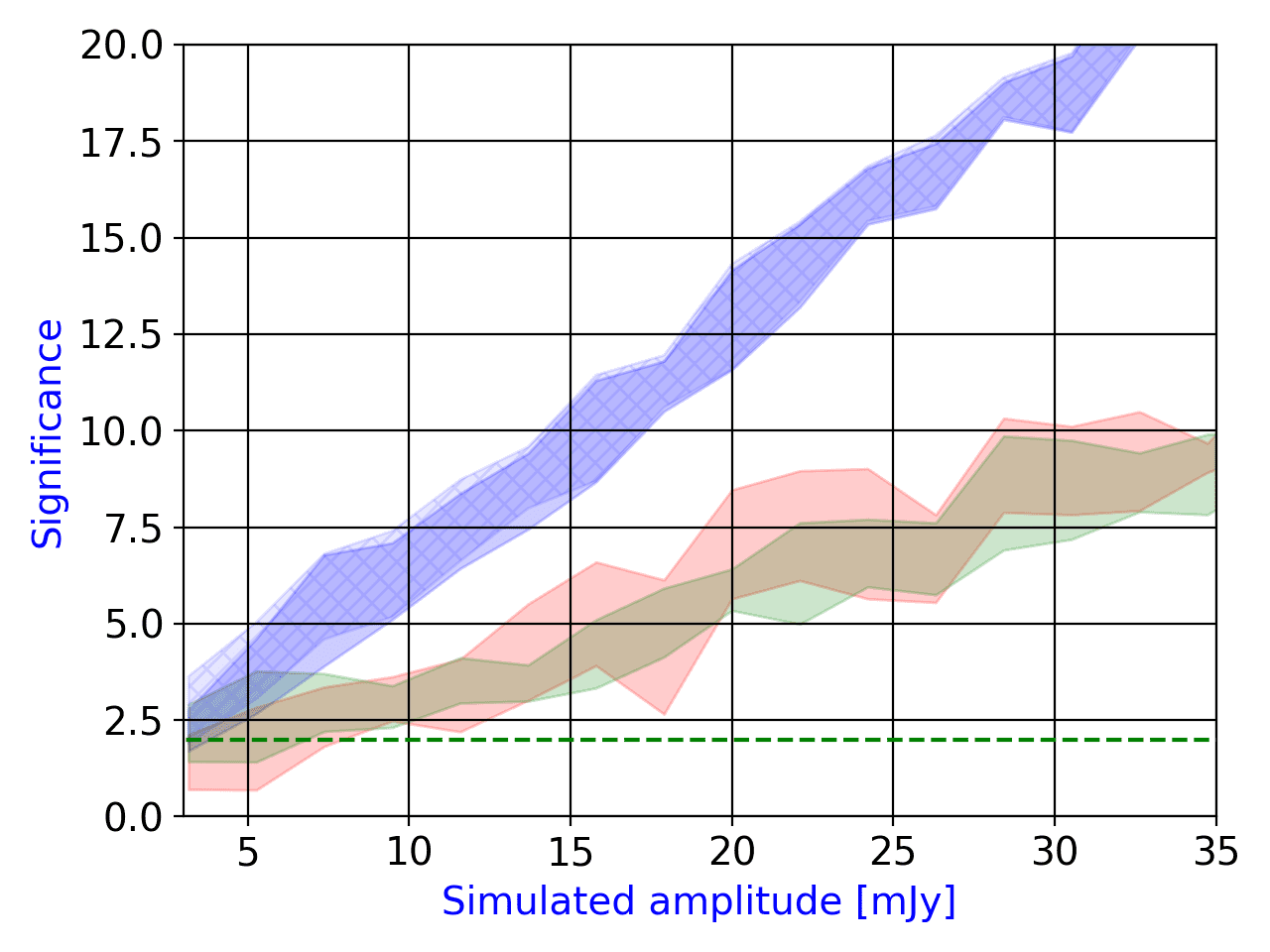}} \quad
{\includegraphics[width=55mm]{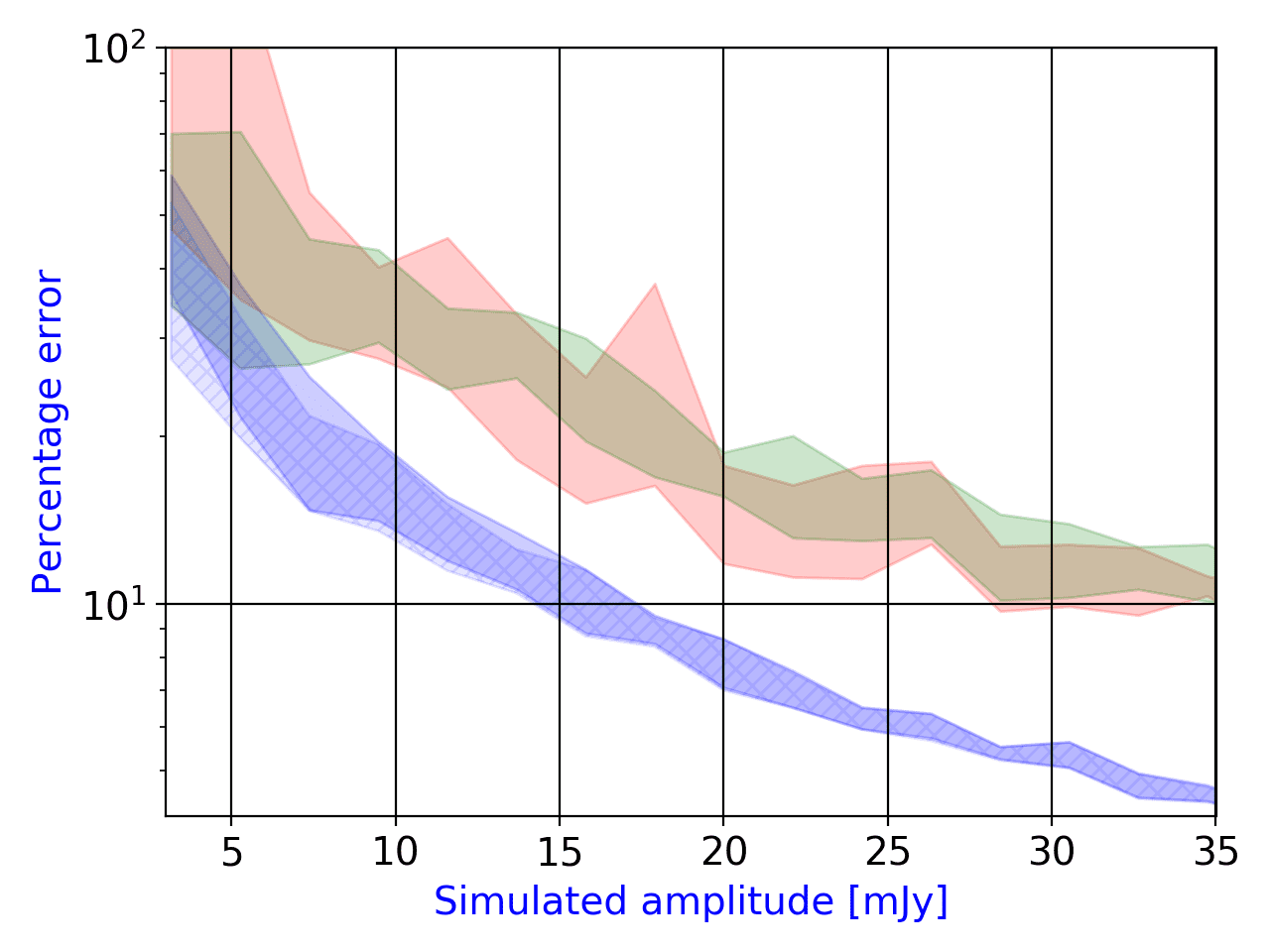}} \\
\caption{Overview of fitting procedure for a fake source (suited by a 2D-Gaussian) about SRT observation of GRB\,190114C at 2019 January 23.
See the caption of Fig.~\ref{fig:sim_tot_rms1} for a full description of the symbols and plots.}
\label{fig:20190123}
\end{figure*}

The third epoch at 2019 January 23 (Fig.~\ref{fig:20190123}) is characterized by $RMS_{min} \sim 2.7$~mJy; Table~\ref{tab:minrms}), resulting in a standard upper limit estimation of $5.4$~mJy ($2\sigma$-level).
The analysis on this epoch seems to confirm the results for the full simulation procedure.
The most accurate method is the method C, and the worst method is the method A (Fig.~\ref{fig:20190123}); moreover, methods A overestimates the flux densities, whereas method B underestimates these flux densities (Fig.~\ref{fig:20190123}).
The injection of fake source in this field shows suggests a $2\sigma$-level upper limit at $\sim 6.5$~mJy for Method A, $\sim 5.1$~mJy for Method B and $\sim 3.0$~mJy for Method C (Fig.~\ref{fig:20190123}); the additional criterion for the free position case for Method C shows a detection at $\sim 6.5$~mJy, where the significance $\mathcal{S}$ is $\sim 4$ ($\sim 2$ for Method A and $\sim 2.5$ for Method B; Fig.~\ref{fig:20190123}).

\begin{figure*} 
\centering
{\includegraphics[width=55mm]{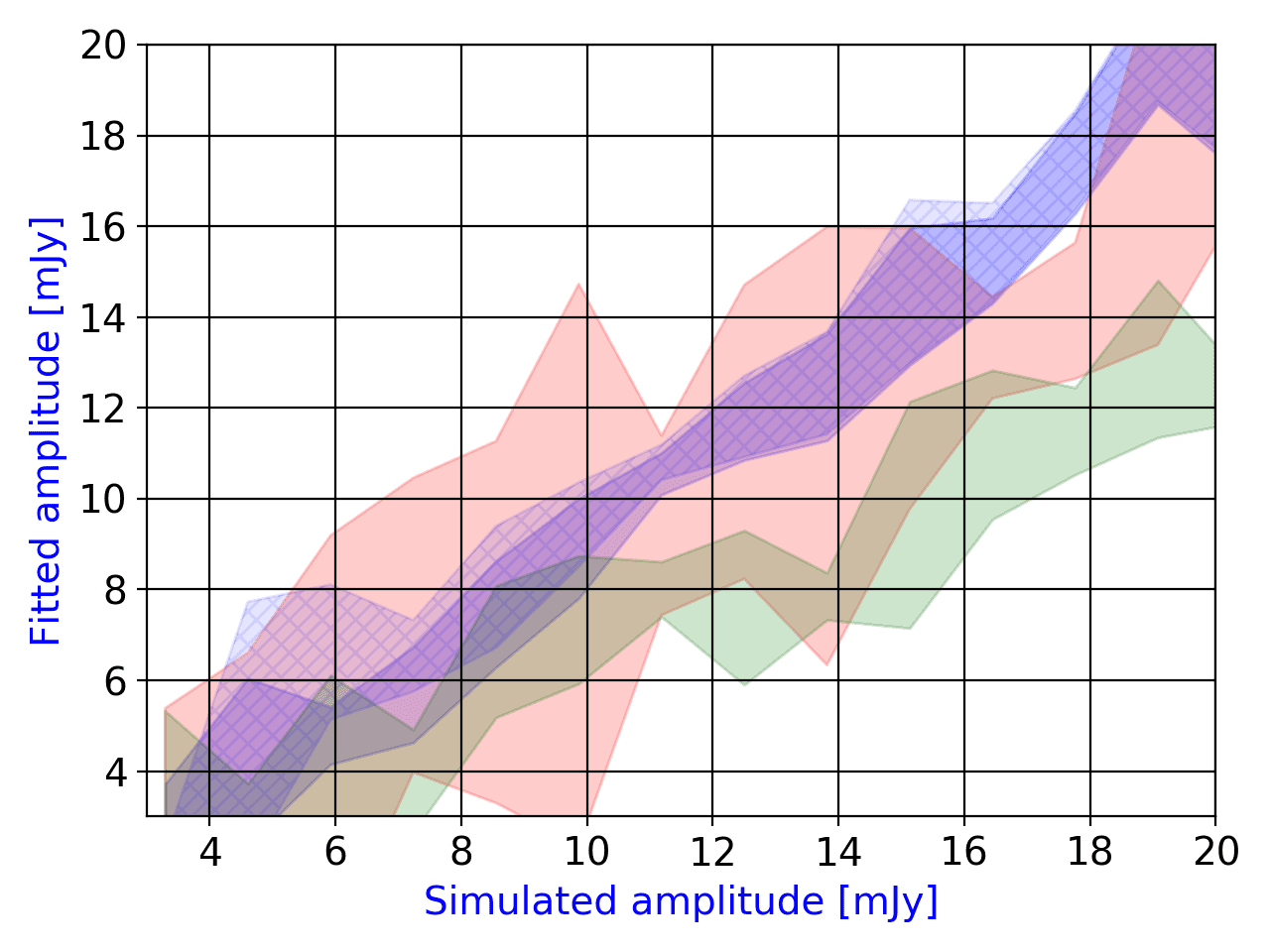}} \quad
{\includegraphics[width=55mm]{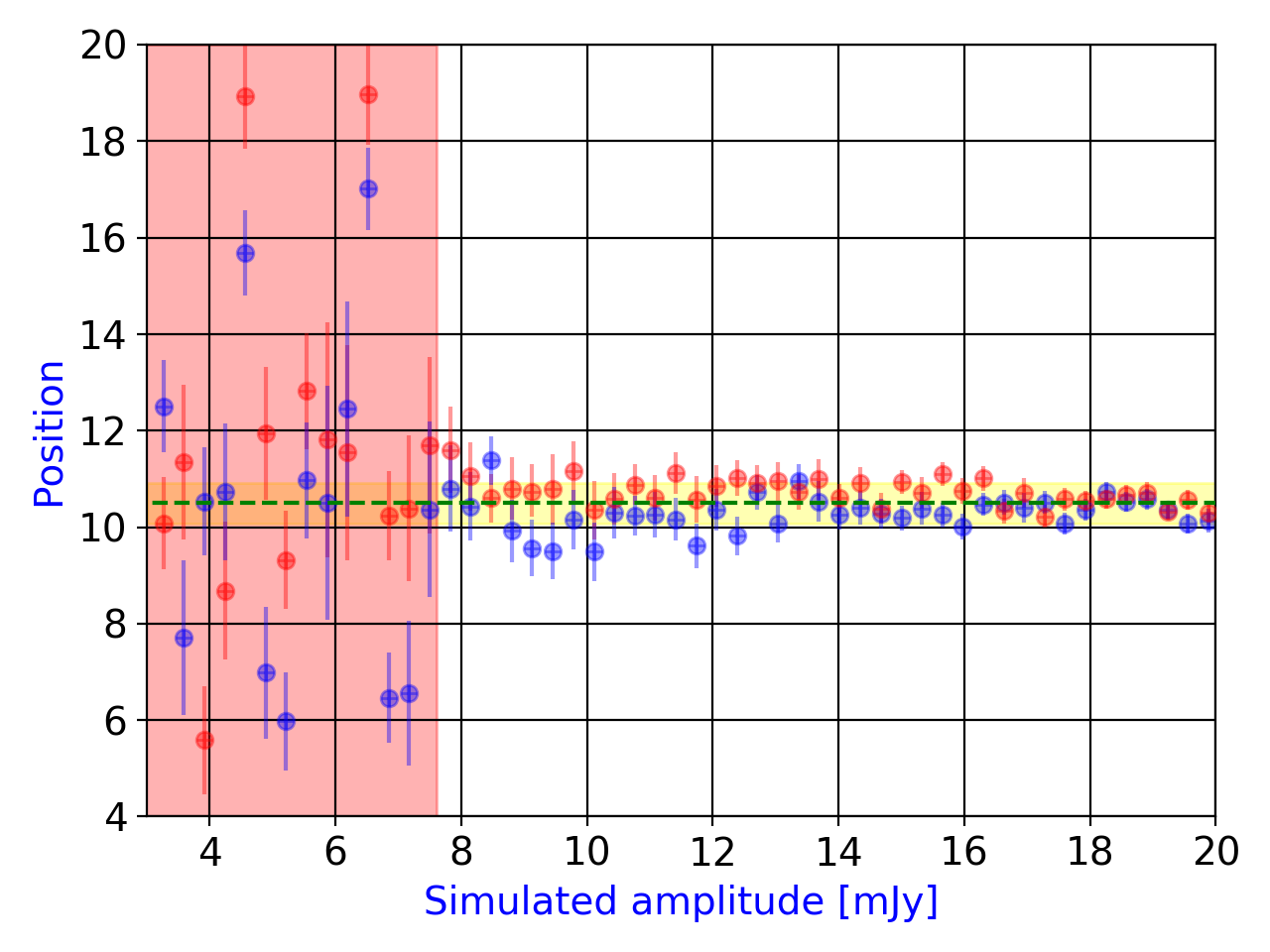}} \\
{\includegraphics[width=55mm]{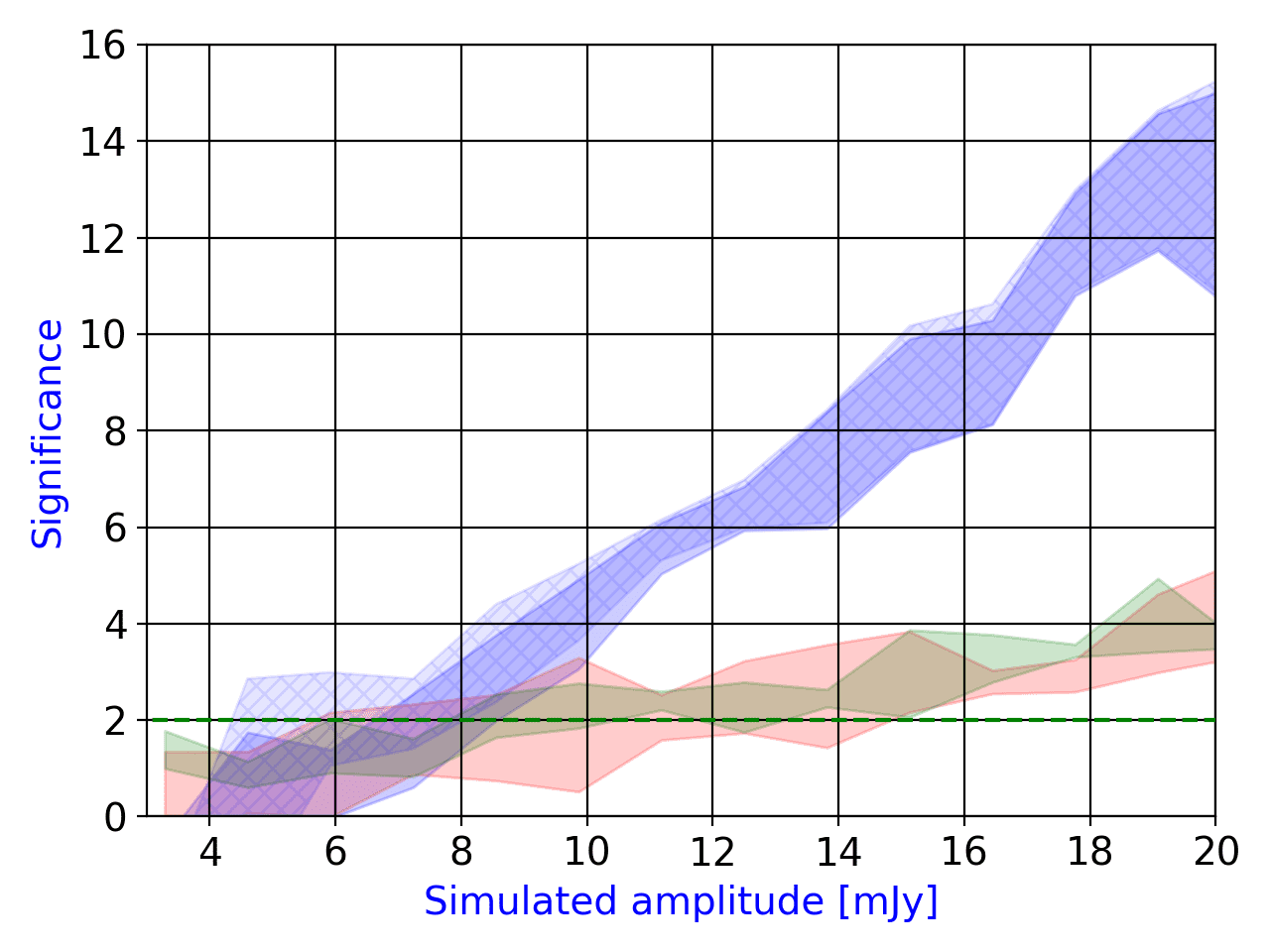}} \quad
{\includegraphics[width=55mm]{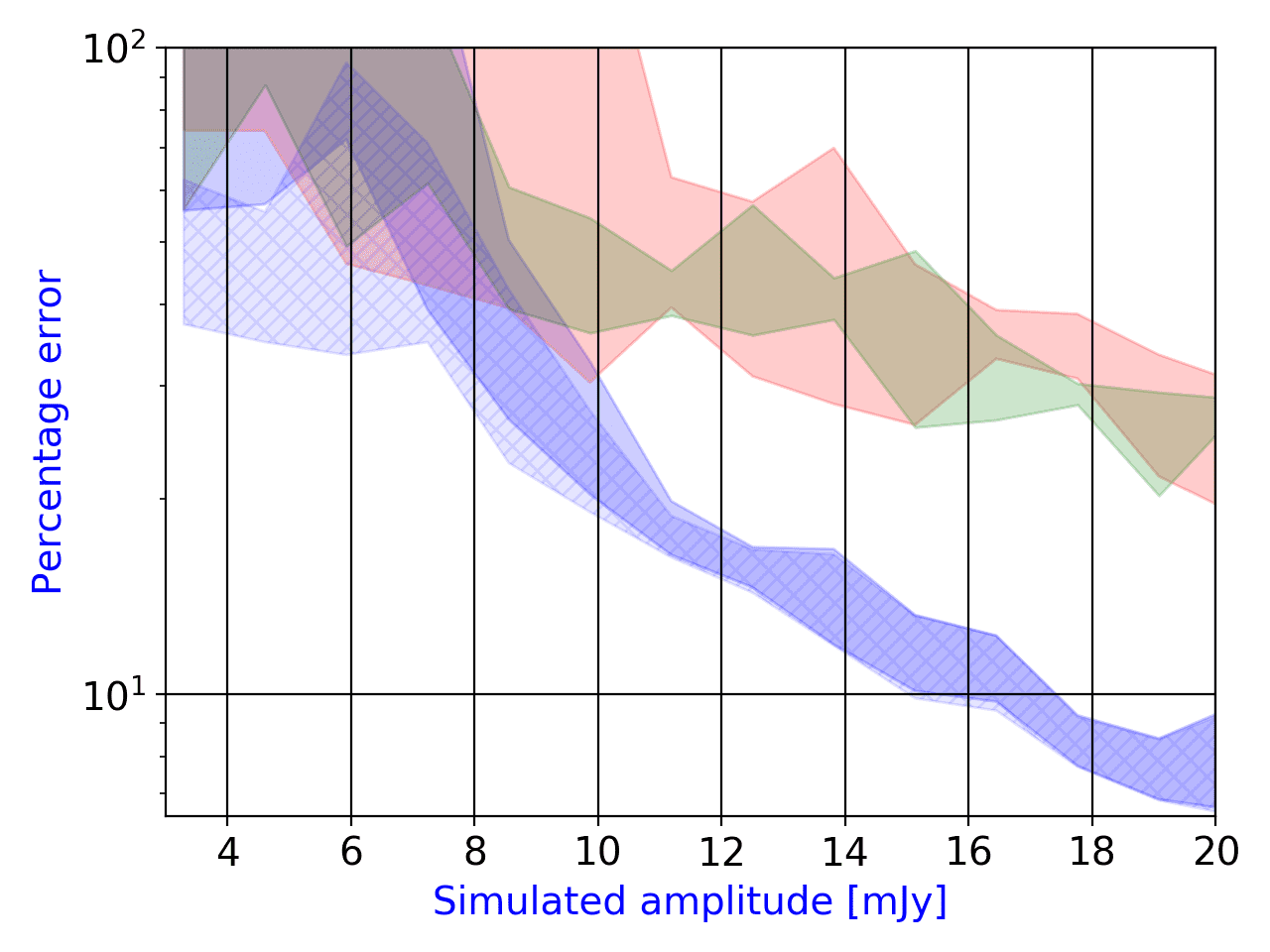}} \\
\caption{Overview of fitting procedure for a fake source (suited by a 2D-Gaussian) about SRT observation of GRB\,181201A at 2019 January 30.
See the caption of Fig.~\ref{fig:sim_tot_rms1} for a full description of the symbols and plots.}
\label{fig:20190130}
\end{figure*}

The fourth epoch at 2019 January 30 shows $RMS_{min} \sim 3.3$~mJy (Table~\ref{tab:minrms}).
As we can see in Fig.~\ref{fig:20190130}, the most accurate method is the method C, and the worst method is the method A.
The injection of fake source in this field suggests a $2\sigma$ upper limit at $\sim 11$~mJy for Method A, $\sim 8.5$~mJy for Method B and $\sim 7.5$~mJy for Method C (Fig.~\ref{fig:20190130}); the additional criterion for Method C (free position case with a fake source) shows a detection at $\sim 7.6$~mJy, where the significance $\mathcal{S}$ is $\sim 3$ ($\sim 2$ and $\sim 1.5$ for Method B and A, respectively; Fig.~\ref{fig:20190130}).

\begin{figure*} 
\centering
{\includegraphics[width=55mm]{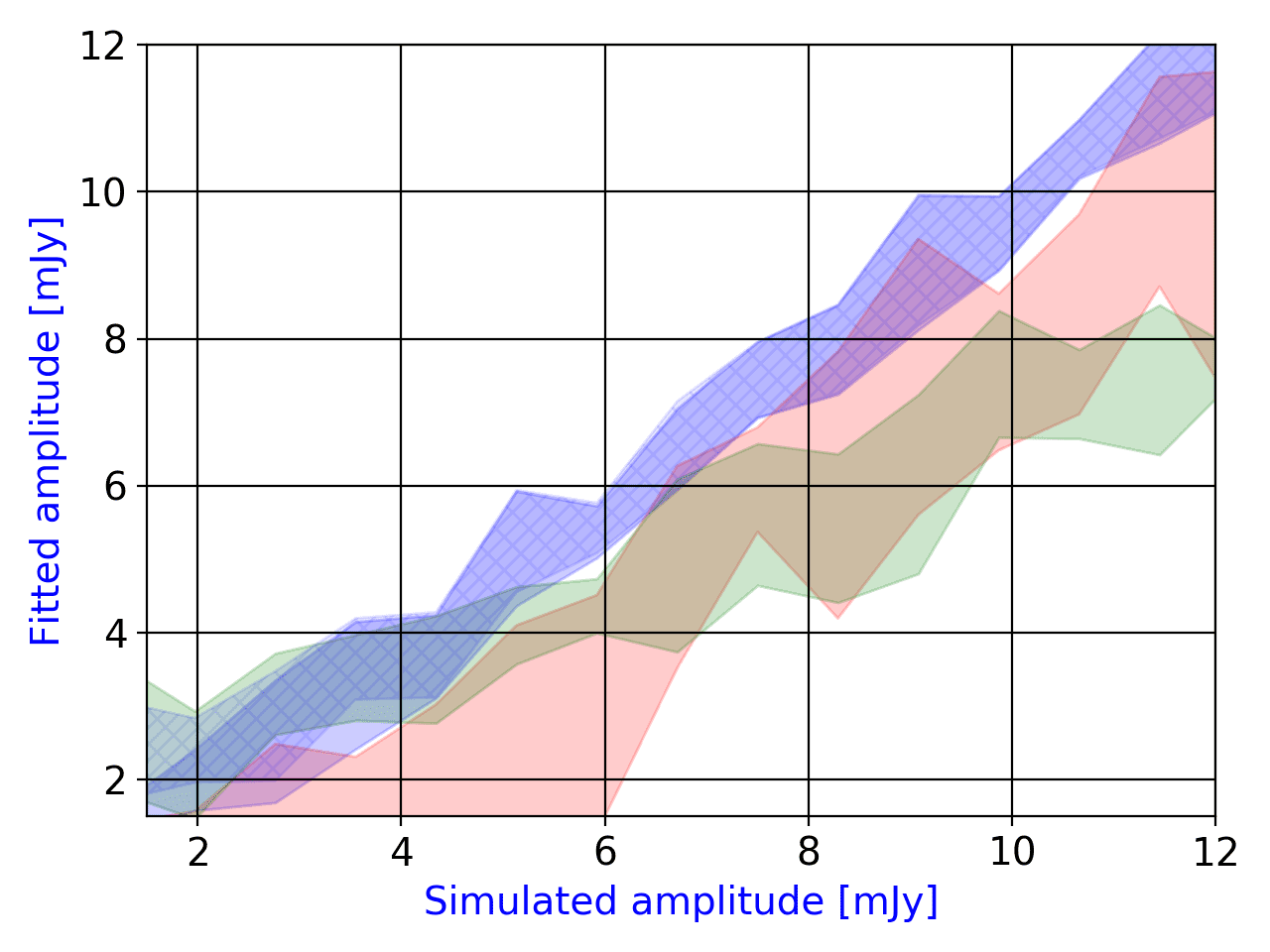}} \quad
{\includegraphics[width=55mm]{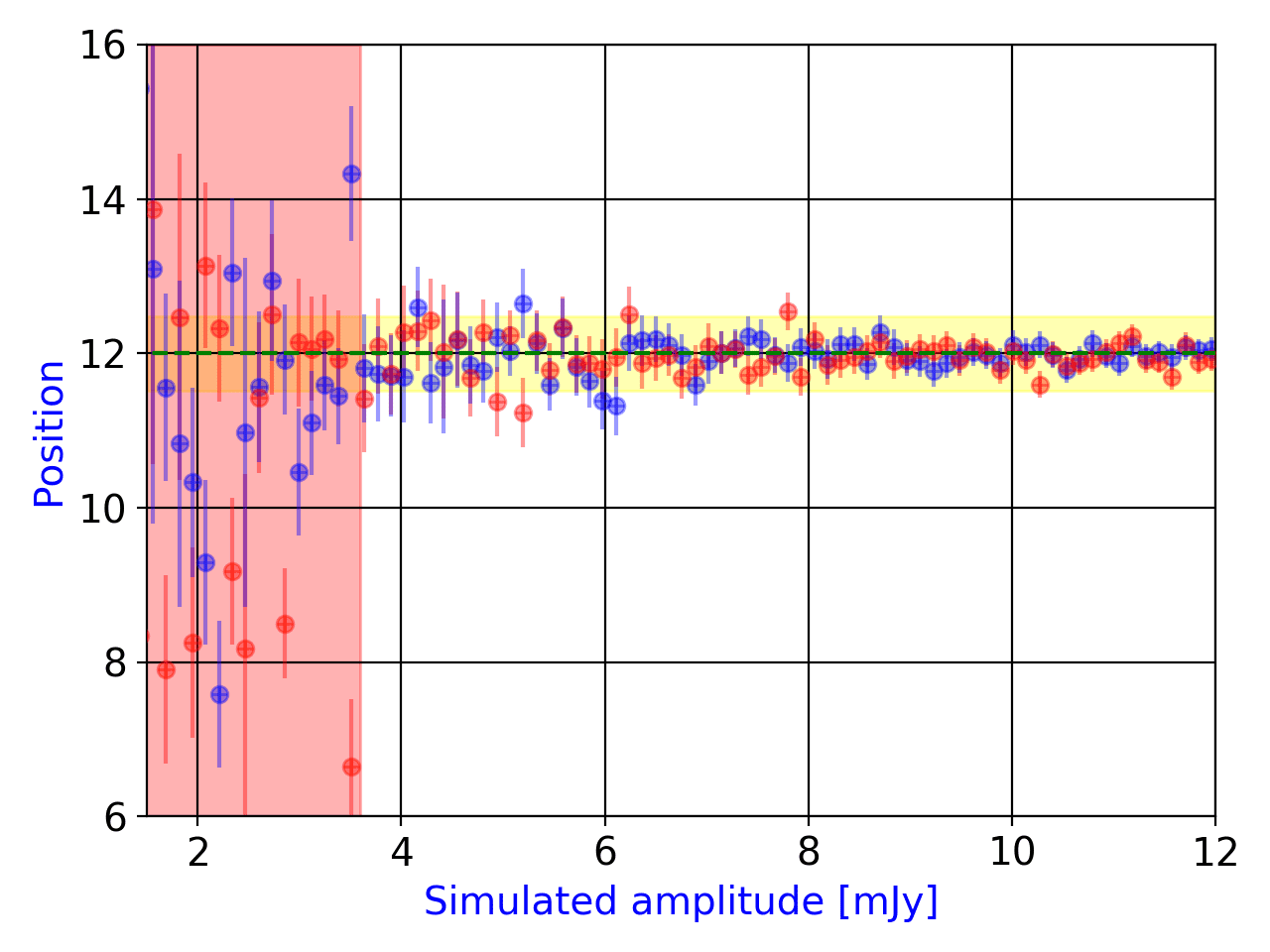}} \\
{\includegraphics[width=55mm]{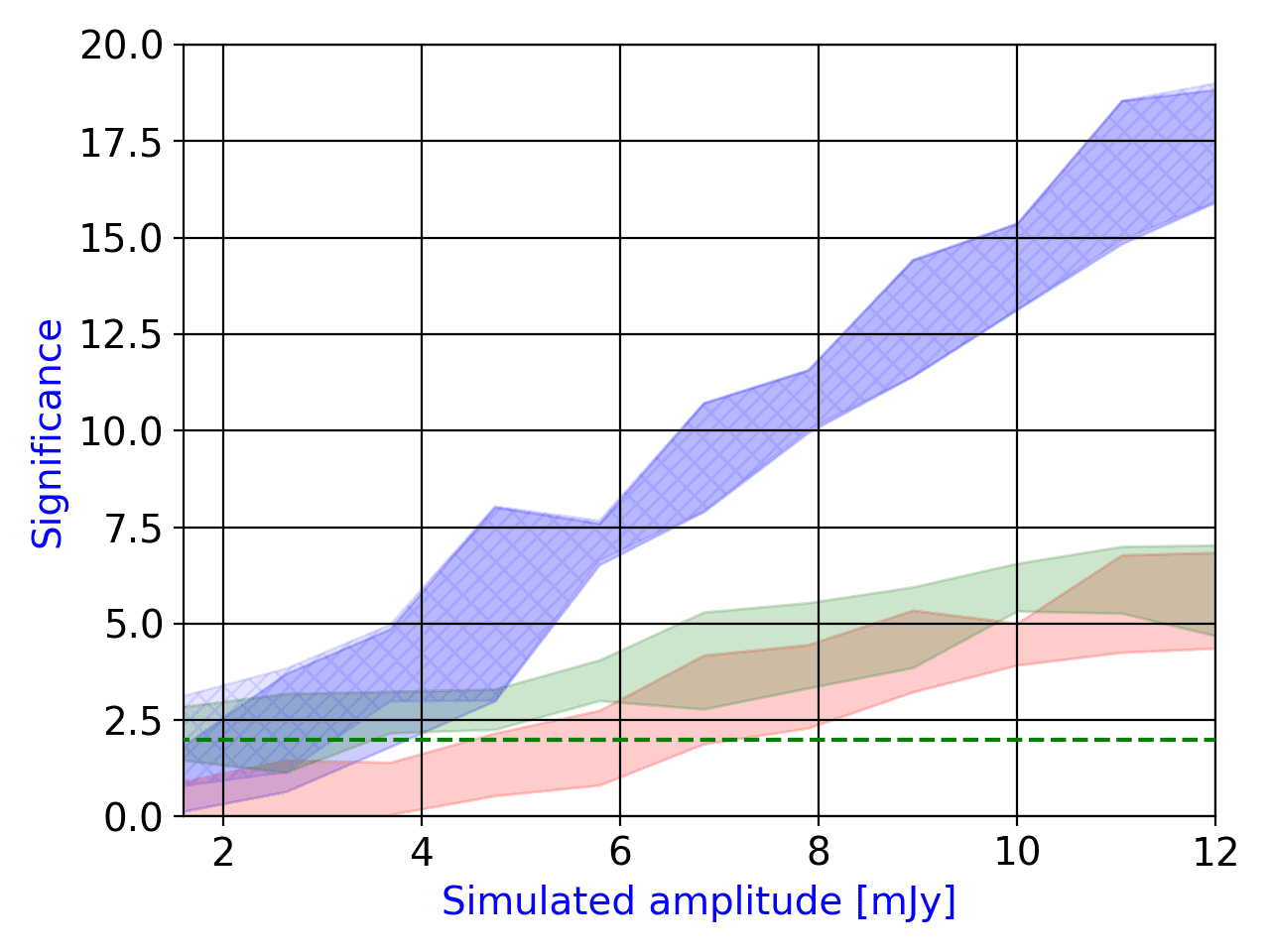}} \quad
{\includegraphics[width=55mm]{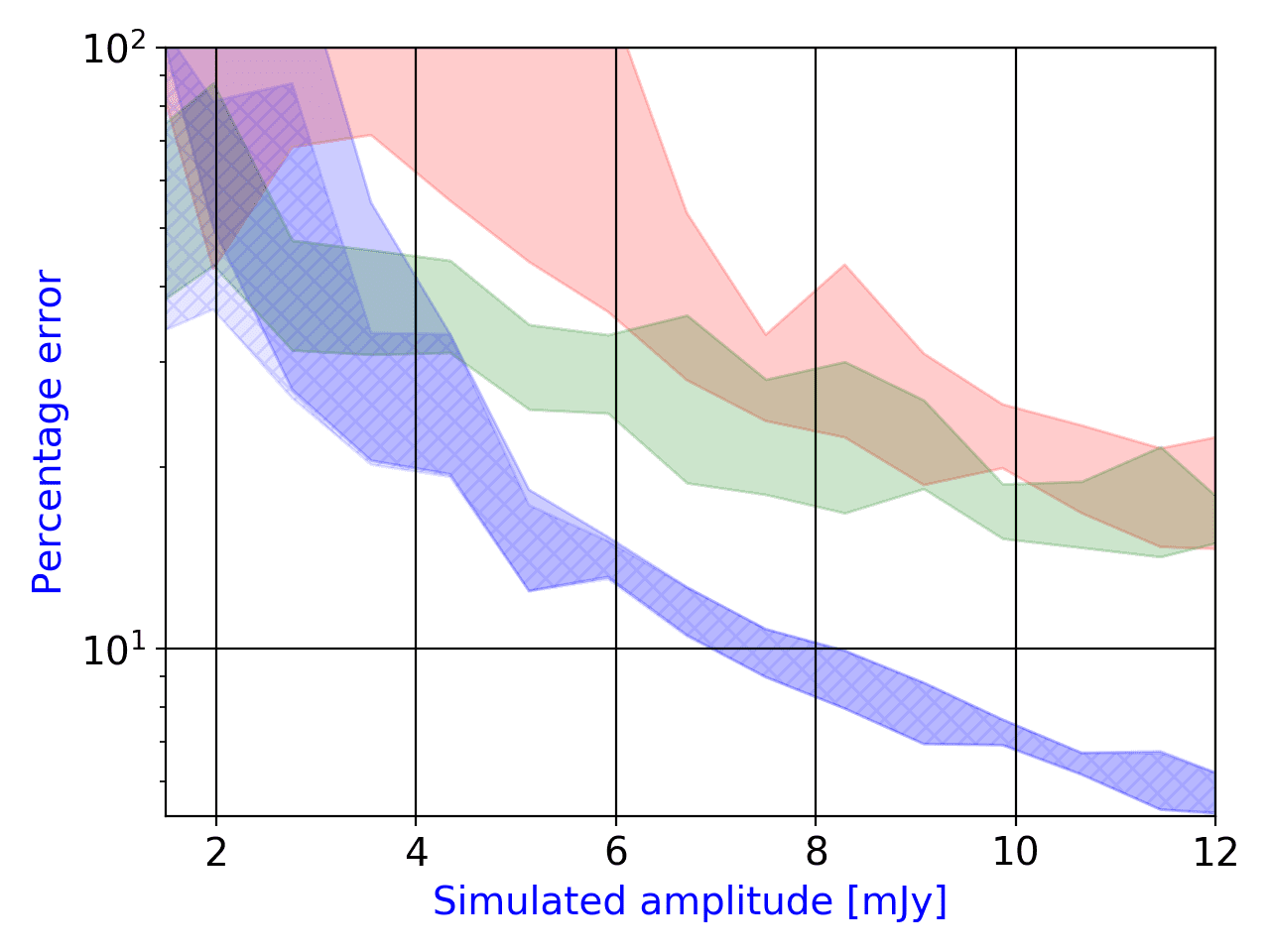}} \\
\caption{Overview of fitting procedure for a fake source (suited by a 2D-Gaussian) about SRT observation of GRB\,190114C at 2019 March 5.
See the caption of Fig.~\ref{fig:sim_tot_rms1} for a full description of the symbols and plots.}
\label{fig:20190305}
\end{figure*}

The fifth epoch at 2019 March 5 shows the lowest value of $RMS_{min}$ in our analysis ($\sim 1.3$~mJy, Table~\ref{tab:minrms}).
The analysis, as in the previous cases, confirms the results for the full simulation procedure (Sect.~\ref{sec:totsim}), where the most accurate method is the method C, and the worst method is the method A (Fig.~\ref{fig:20190305}).
The injection of fake source in this field suggests a $2\sigma$-level upper limit at $\sim 5.2$~mJy for Method A, $\sim 2.4$~mJy for Method B and $\sim 2.5$~mJy for Method C (Fig.~\ref{fig:20190305}); the additional criterion for the free position case for Method C shows a detection at $\sim 3.5$~mJy, where the significance $\mathcal{S}$ is $\sim 4$ ($\sim 3$ and $\sim 1$ for Method B and A, respectively; Fig.~\ref{fig:20190305}).
This epoch shows the power of the Method C in terms of accuracy and robustness: at the same conditions we are able to detect a source where Method A fails (Fig.~\ref{fig:20190305}, bottom right).

\begin{figure*} 
\centering
{\includegraphics[width=55mm]{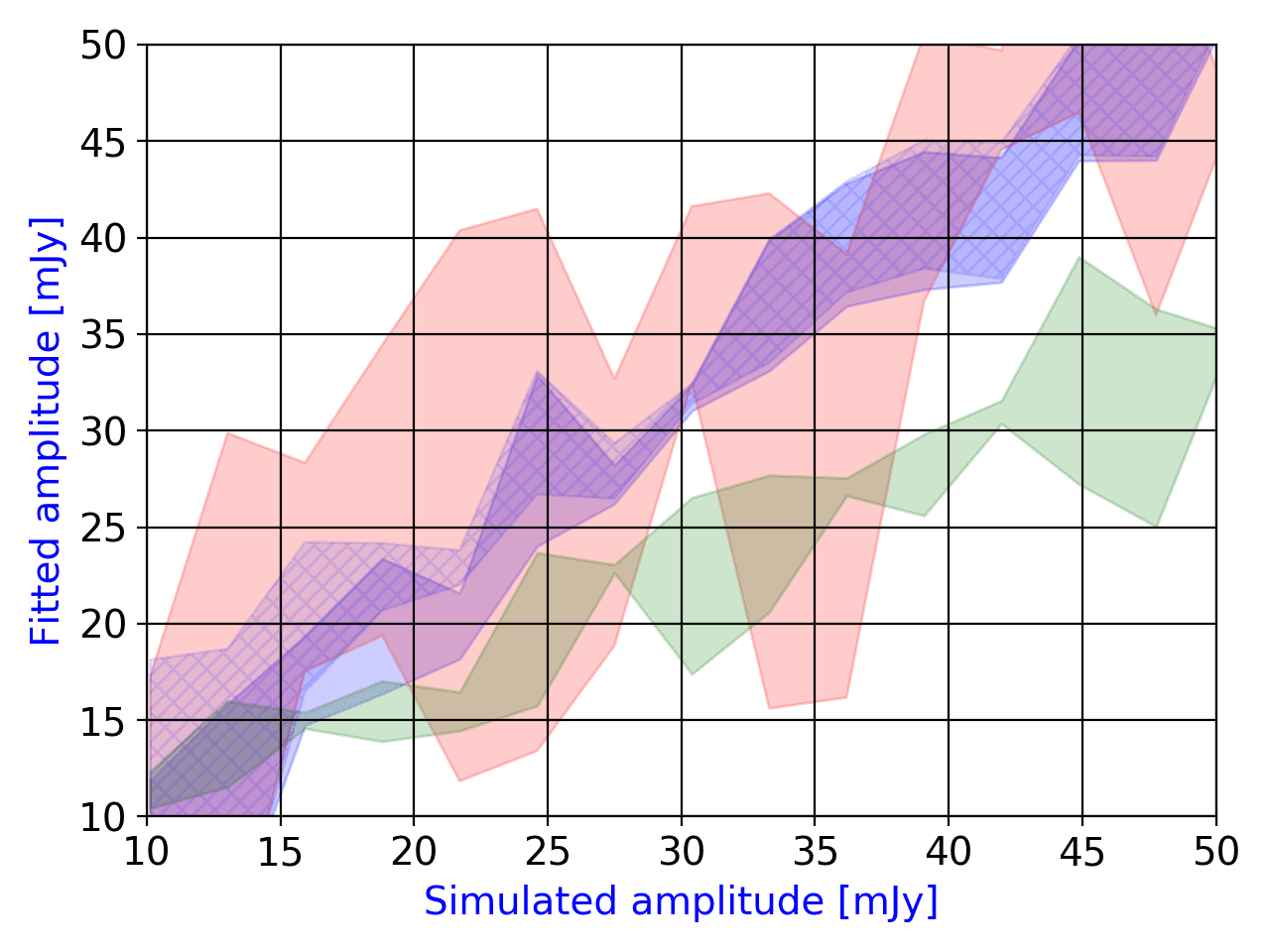}} \quad
{\includegraphics[width=55mm]{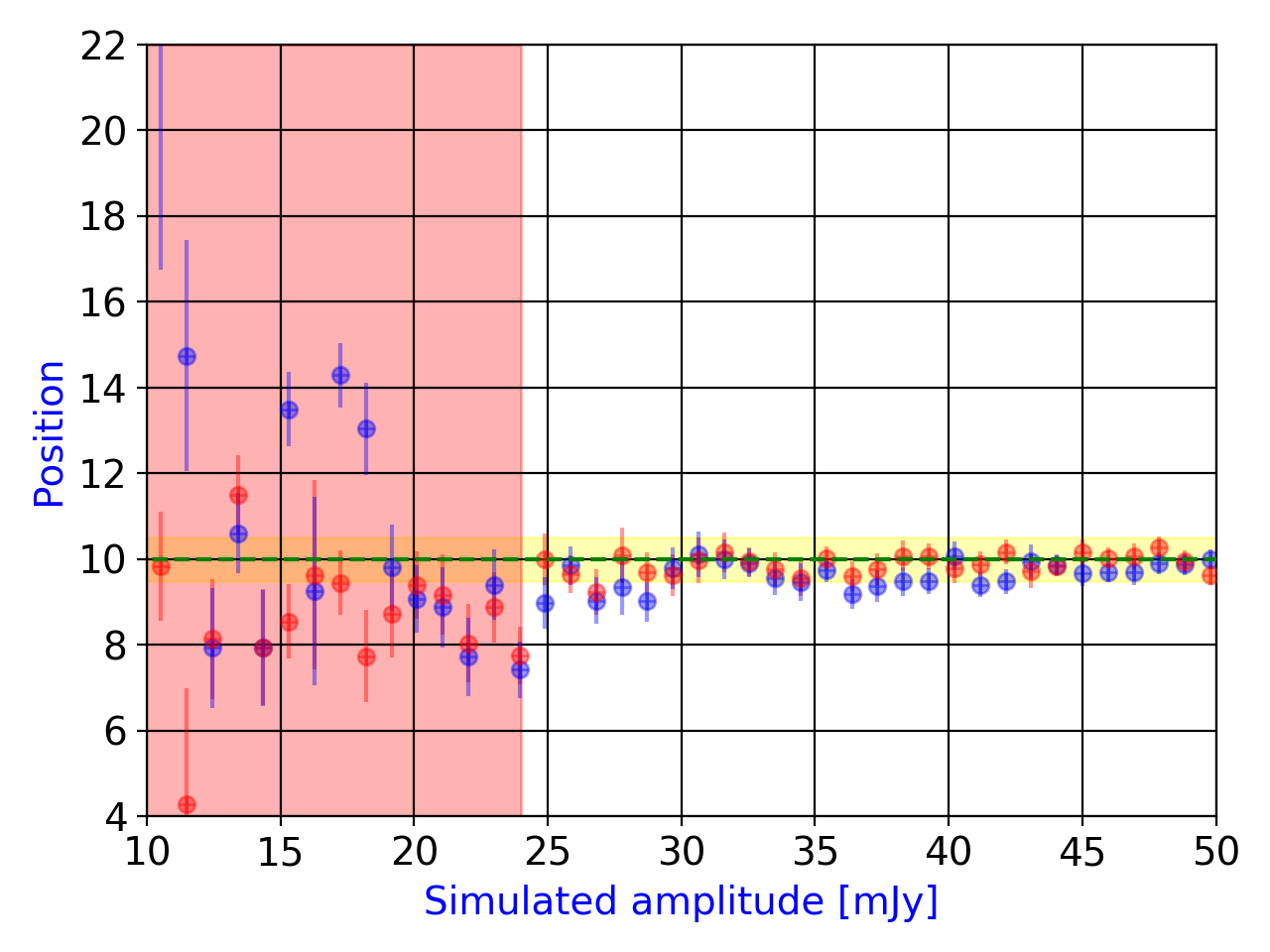}} \\
{\includegraphics[width=55mm]{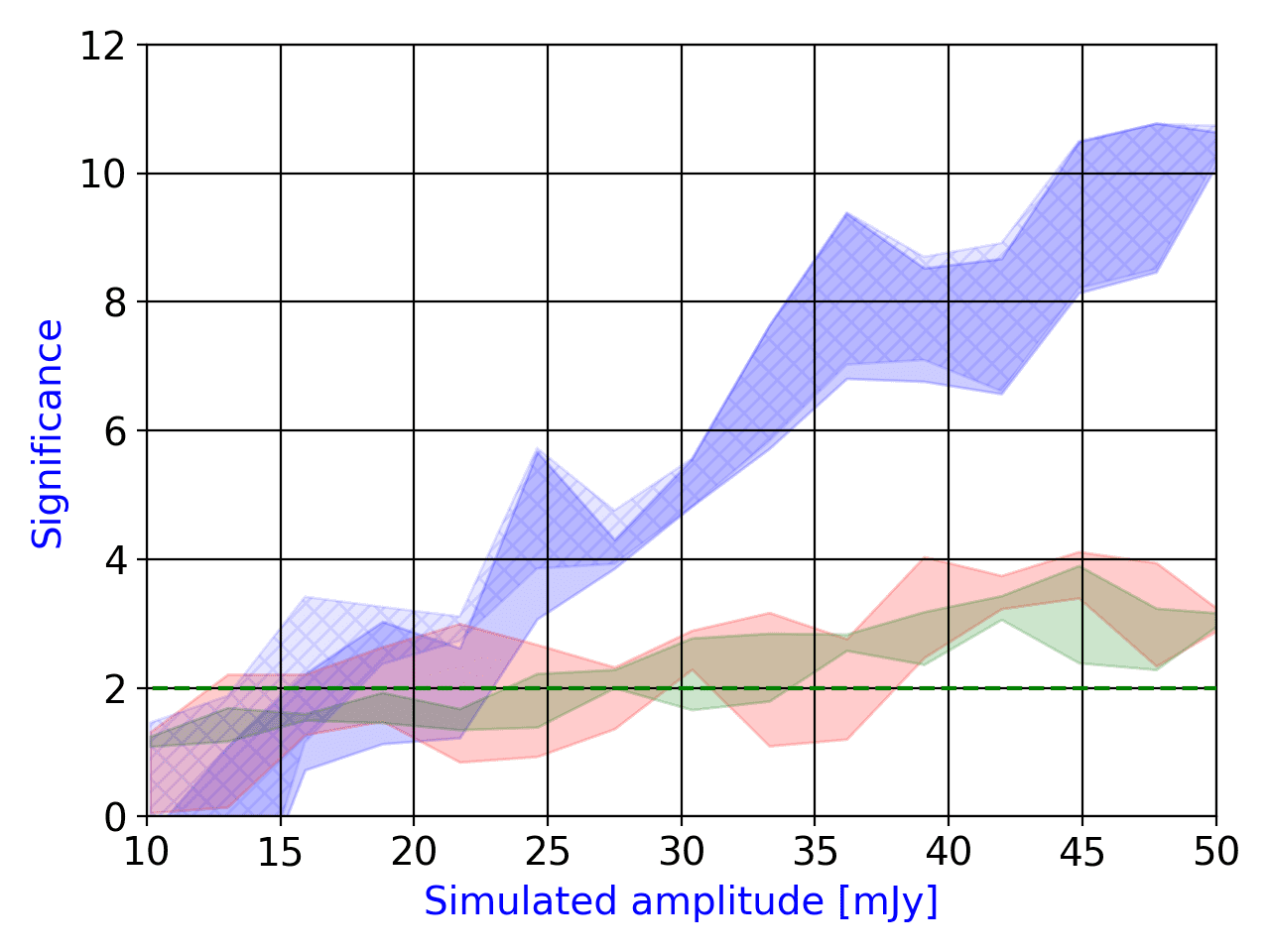}} \quad
{\includegraphics[width=55mm]{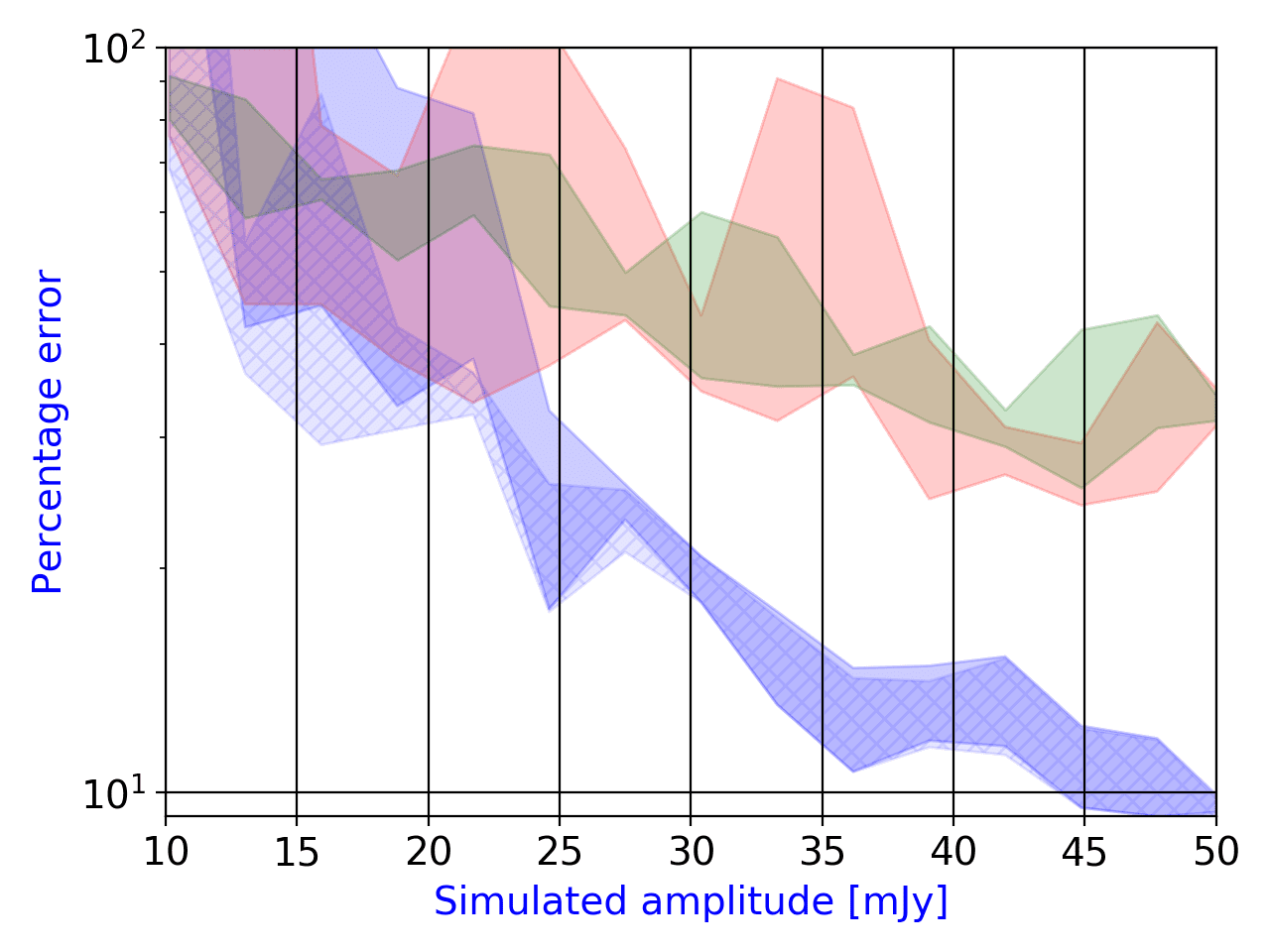}} \\
\caption{Overview of fitting procedure for a fake source (suited by a 2D-Gaussian) about SRT observation of GRB\,181201A at 2019 March 22.
See the caption of Fig.~\ref{fig:sim_tot_rms1} for a full description of the symbols and plots.}
\label{fig:20190322}
\end{figure*}

On the other hand, the last epoch at 2019 March 22 is the worst image in our followup campaign, characterized by a very bad weather at SRT site (rain and wind) and $RMS_{min} \sim 9.6$~mJy (Table~\ref{tab:minrms}).
The injection of fake source in this field suggests a $2\sigma$-level upper limit at $\sim 36.4$~mJy for Method A, $\sim 25.9$~mJy for Method B and $\sim 18.2$~mJy for Method C (Fig.~\ref{fig:20190322}).
The additional criterion for the free position case for Method C shows a detection at $\sim 24$~mJy, where the significance $\mathcal{S}$ is $\sim 4$ ($\sim 2$ for Method A and B; Fig.~\ref{fig:20190322}).

\section{Discussion}
\label{sec:discussion}

Radio observations of real fields are intrinsically characterized by several features (such as the background, systematic errors of the radio devices or weather conditions) which directly impact on the detection and flux estimation of -- especially faint -- sources.
The analysis of real images obtained with SDI confirms the results of the full simulation procedure (Sect.~\ref{sec:totsim}), showing that -- especially for faint sources (mJy or sub-mJy events) -- a refined knowledge of the baseline (such as the background, other sources, RFI, weather, and the systematic errors of the antenna) is crucial.
These images show that Method C is the most accurate, whereas the least accurate one is Method A, which either under- or overestimates flux densities, possibly due to the approximation of the 2D-Gaussian peak with respect to the corresponding pixel peak.

Method B tends to underestimate flux densities, possibly due to inaccurate baseline subtraction in SDI package, considering that this method does not consider RFI or other sources near the target.
The baseline subtraction in SDI package is crucial especially for faint sources. In this respect, Method C is also particularly sensitive to the baseline subtraction.

In Section~\ref{sec:realmaps} we estimated $2\sigma$ upper limits and set strong positional constraints, emphasizing how the knowledge of background is essential. These limits strongly depend on the weather conditions (e.g. rain, humidity, and cloud cover), especially for observations of very faint sources, as is our case.
From the analysis of real images, we do not detect the afterglows of either 181201A or 190114C in any images at $\mathcal{S} \gtrsim 5$.
Our upper limits lie above the VLA and ATCA flux densities \cite{Laskar19b,Misra19} and East Asia VLBI upper limits \cite{Tao20} obtained almost at the same epoch.
In particular, our upper limit of 17 January 2019 for 190114C ($1.8$~mJy), estimated with Method C in fixed position case (Table~\ref{tab:quickand2}), lies just above the interpolated ATCA flux density ($\sim 1.6$~mJy) calculated in C-band ($6.9$~GHz) at the time of SRT observation ($2.9$~d after the GRB trigger). This value is affected by scintillation for about $35\%$ of the observed flux density \cite{Misra19}, and hence -- even more so -- is compatible with our upper limit.

In the literature there is an upper limit ($0.6$~mJy at $3\sigma$) with SRT in C-band for an old GRB afterglow (GRB\,151027A, \cite{Nappo17}), significantly lower than our values.
It is not straightforward to compare results obtained under very different observing conditions, so the discrepancy should not necessarily be alarming.
In particular, \cite{Nappo17} did not carry out a mapping, but limited their observations to cross-scans centered on source, following an approach designed for blazar observations \cite{Giroletti20}.
While such a technique benefits from a higher efficiency (through a larger fraction of time spent on-source), it is on the other hand more limited in the presence of a complex background. Mapping thus represents a more reliable and general approach, particularly in the Galactic plane and/or at high frequency.
 
The success of the observations of faint sources (or the quality of the image) for SRT is described by the limiting sensitivity of a radio astronomical receiver, calculated with the radiometer equation $S = \phi \Delta t^{-1/2}$, where $S = RMS_{min}$ (in units of mJy) and $\Delta t$ is the total observing time of the source (in units of hours).
Therefore we calculate the parameter $\phi = S \Delta t^{1/2}$ (in units of mJy~h$^{1/2}$) that provides a rough estimation of all the contamination factors of the image, such as the background, other sources beyond our target, RFI phenomenon, weather conditions, and systematic errors of the single-dish facility.
Our radio campaign shows that (1) $\phi \leq 3.5$~mJy~h$^{1/2}$ indicates a high-quality image (with a good upper limit), (2) $\phi$ between $3.5$ and $8$~mJy~h$^{1/2}$ indicates a medium-quality image, and (3) $\phi > 8$~mJy~h$^{1/2}$ indicates a low-quality image (Table~\ref{tab:minrms}).
\begin{table*}
\centering
\caption{Minimum value of $RMS_{min}$ (estimated with Method A) for the undetected sources observed in SRT maps (in units of mJy), based on $1.5$~beams of SRT receiver.
$\delta t$ indicates the integration time (in units of hours), and $\phi = RMS_{min}/ \delta t$ (in units of mJy h$^{1/2}$). }
\label{tab:minrms}
\begin{tabular}{ccc|ccc}
\hline
Epoch & GRB & Receiver & $\delta t$ & $RMS_{min}$ & $\phi$ \\
(aaaa/mm/dd) & & & (h) & (mJy) & (mJy h$^{1/2}$) \\
\hline
2018/12/11 & 181201A & SRT-C & $3.84$ & $1.93$ & $3.78$  \\
2019/01/17 & 190114C & SRT-C & $4.80$ & $1.49$ & $3.26$  \\
2019/01/23 & 190114C & SRT-C & $4.80$ & $2.70$ & $5.92$  \\
2019/01/30 & 181201A & SRT-C & $5.28$ & $3.26$ & $7.49$  \\
2019/03/05 & 190114C & SRT-C & $3.60$ & $1.30$ & $2.47$  \\
2019/03/22 & 181201A & SRT-C & $4.08$ & $9.57$ & $19.33$ \\
\hline \hline
\end{tabular}
\end{table*}

\section{Conclusions}
\label{sec:conc}

We analyzed three detection methods having different degrees of sensitivity and robustness (`quick-look' or Method A, `source extraction' or Method B, fitting procedure with a 2-D Gaussian or Method C).
Their performances were assessed in the case of weak sources in single-dish mode through the INAF network of radio telescopes.
To this aim, we developed a specific Python code, where the input data are the calibrated (in units of flux density/beam) FITS images (suited for INAF network) produced by SDI, and the output consists in flux densities of the target and corresponding uncertainties.

This new approach for the SRT data analysis enhances the capabilities of this radio telescope, especially optimizing the detection of faint sources, as for GRB afterglows or GW radio counterparts.
We observed two GRB afterglows (181201A and 190114C) and the Galactic binary  GRS\,1915+105 with SRT in C-band ($6.9$~GHz).

Our comparative analysis of the different detection methods made extensive use of simulations as a useful complement of actual radio observations.
In the regime of faint/undetected sources, simulations of injected point-like sources are used for the assessment; in particular, the estimated flux densities (or upper limits in case of undetected sources) strongly depend on the weather conditions (e.g. rain, humidity, and cloud cover).
Simulations of both background and source are essential to characterize the detection of point-like sources in images/fields, and as such they were used to calibrate our software for the analysis of real targets.
This analysis shows that the Method C provides an excellent agreement between fitted and real injected flux density.
Source detection at $3\sigma$ confidence (for $N_0 \gtrsim 1$~mJy) is feasible for flux densities $\gtrsim 2$~mJy; on the other hand, Method A shows high uncertainties and we recommended it only for a rapid and preliminary estimation of the flux density and/or when the instrument beam is poorly known.

These results are further corroborated by the analysis of real radio observations.
For GRS\,1915+105, Method C (in free position case) is able to detect it, although the region is characterized by a variable background.
Images with undetected sources offer the possibility to study the conditions required for a detection through the injection of a fake source in the real radio image, located at the position previously found by other facilities.
Our results show that -- especially for faint sources (mJy or sub-mJy events) -- a deep knowledge of the radio background is crucial for accurate flux density measures.
These images show that Method C pushes down the sensitivity limits of this radio telescope -- with respect to more traditional techniques -- to $\sim 1.8$~mJy, improving by $\sim 40$\% compared with the initial value.

The image quality for faint sources is described by $\phi$, that provides a rough estimation of all the contamination factors affecting the image; our campaigns show that the range $\phi \leq 3.5$~mJy~h$^{1/2}$ corresponds to a high-quality image, $3.5\le\phi/({\rm mJy~h}^{1/2})\le8$ corresponds to a medium-quality image, whereas $\phi > 8$~mJy~h$^{1/2}$ characterizes a low-quality image.

The code developed for this analysis -- directly implemented in the SDI package in the near future -- will be further improved adopting a Bayesian approach in the fitting procedure that incorporates the likelihood analysis for crowded images.
In the multi-messenger era, it can be employed to analyze (1) not only faint sources such as the radio counterparts of GRBs or GWs, but also other sources such as solar flares \footnote{\url{https://sites.google.com/inaf.it/sundish/home}}, and (2) images obtained through interferometric radio telescopes, such as the Very Large Baseline Array (VLBA\footnote{\url{https://science.nrao.edu/facilities/vlba}}), the European Very Large Baseline Interferometry Network (EVN\footnote{\url{https://www.evlbi.org}}, of which SRT is part), LOw Frequency ARray (LOFAR, \cite{LOFAR13}) or the next generation Square Kilometer Array facility (SKA, \cite{ASKAP}).

\begin{acknowledgements}
The Sardinia Radio Telescope is funded by the Department of University and Research (MIUR), the Italian Space Agency (ASI), and the Autonomous Region of Sardinia (RAS), and is operated as a National Facility by the National Institute for Astrophysics (INAF).
M.~Marongiu gratefully acknowledges the University of Ferrara for the financial support of his PhD scholarship.
We acknowledge the support from PRIN-INAF-2016 ``Towards the SKA and CTA era: the discovery, localization, and physics of transient sources''.
M.~Marongiu thanks P.~Bergamini and G.~Angora for the useful discussion about Python programming language and data analysis; moreover, M.~Marongiu is very grateful to K.~D.~Alexander for useful conversations.
We acknowledge the TAC, the scheduler and the SRT staff for approving and executing these followup campaigns.

This is a pre-print of an article published in Experimental Astronomy.
The final authenticated version is available online at: \url{https://doi.org/10.1007/s10686-020-09658-9}.
\end{acknowledgements}

\section*{ORCID iDs}

M.~Marongiu: \url{https://orcid.org/0000-0002-5817-4009} \\
A.~Pellizzoni: \url{https://orcid.org/0000-0002-4590-0040} \\
E.~Egron: \url{https://orcid.org/0000-0002-1532-4142} \\
T.~Laskar: \url{https://orcid.org/0000-0003-1792-2338} \\
M.~Giroletti: \url{https://orcid.org/0000-0002-8657-8852} \\
S.~Loru: \url{https://orcid.org/0000-0001-5126-1719} \\
A.~Melis: \url{https://orcid.org/0000-0002-6558-1315} \\
G.~Carboni: \url{https://orcid.org/0000-0002-2056-7501} \\
C.~Guidorzi: \url{https://orcid.org/0000-0001-6869-0835} \\
S.~Kobayashi: \url{https://orcid.org/0000-0001-7946-4200} \\
N.~Jordana-Mitjans: \url{https://orcid.org/0000-0002-5467-8277} \\
A.~Rossi: \url{https://orcid.org/0000-0002-8860-6538} \\
C.~G.~Mundell: \url{https://orcid.org/0000-0003-2809-8743} \\
R.~Concu: \url{https://orcid.org/0000-0003-3621-349X} \\
R.~Martone: \url{https://orcid.org/0000-0002-0335-319X} \\
L.~Nicastro: \url{https://orcid.org/0000-0001-8534-6788} \\

\section*{Conflict of interest}
The authors declare that they have no conflict of interest.

\bibliographystyle{spphys}          
\bibliography{biblio}               

\end{document}